\documentclass{aa}
\usepackage{booktabs}
\usepackage{mathtools}
\usepackage{txfonts}
\usepackage{natbib}
\usepackage{graphicx}
\usepackage{epstopdf}
\usepackage{pgf}
\usepackage{soul}
\definecolor{lightgreen}{HTML}{B7F774}
\definecolor{lightred}{HTML}{FF6666}
\definecolor{lightorange}{HTML}{FE9A2E}
\sethlcolor{lightgreen}

\usepackage[normalem]{ulem}

\providecommand{\abs}[1]{\lvert#1\rvert}
\providecommand{\norm}[1]{\lVert#1\rVert}
\DeclareMathOperator{\dilation}{dilation}

\begin{document}

\title{Wavelet-based decomposition and analysis of structural patterns
  in astronomical images}

\author{Florent Mertens\inst{1}
        \and
        Andrei Lobanov\inst{1,2}}

    \institute{Max-Planck-Institut f\"ur Radioastronomie,
              Auf dem Hugel 69, 53121 Bonn, Germany
    \and
    Institut f\"ur Experimentalphysik, Universit\"at Hamburg, 
    Luruper Chaussee 149, 22761 Hamburg, Germany}

  \date{}
 
  \abstract   
  {Images of spatially resolved astrophysical objects contain a wealth of
  morphological and dynamical information, and effectively extracting this
  information is of paramount importance for understanding     the physics and
  evolution of these objects. The algorithms and methods     currently employed
  for this purpose (such as    Gaussian model fitting) often use simplified
  approaches to describe the structure of resolved objects.}
  {Automated (unsupervised) methods for structure decomposition and     tracking
  of structural patterns are needed for this purpose to be able to treat the
  complexity of structure and large amounts of data involved.}
  {We developed a new wavelet-based image segmentation and evaluation (WISE)
  method for multiscale decomposition, segmentation, and tracking of structural
  patterns in astronomical images.}
  {The method was tested against simulated images of relativistic jets   and
  applied to data from long-term monitoring of parsec-scale radio   jets in
  3C\,273 and 3C\,120. Working at its coarsest resolution,   WISE reproduces the
  previous results of a model-fitting evaluation of   the structure and kinematics
  in these jets exceptionally well. Extending the WISE structure analysis to
  fine scales provides the first robust measurements of   two-dimensional velocity
  fields in these jets and indicates that the velocity fields probably reflect the
  evolution of Kelvin-Helmholtz instabilities that develop in the flow.}
   {}


   \maketitle
%

\section{Introduction}

The steady improvements of the dynamic range of astronomical images and the
ever-increasing complexity and detail of astrophysical modeling bring a higher
demand on automatic (or {\em \textup{unsupervised}}) methods for characterizing
and analyzing structural patterns in astronomical images.

Several of the approaches developed in the fields of computer vision and remote-
sensing to track structural changes \citep[{\em   cf.}][]{yuan_d_1998,doucet_a_1999,
arulampalam_m_2002,sidenbladh_h_2004,doucet_a_2005,myint_comparison_2008}
either require oversampling in the temporal domain or rely on multiband
(multicolor) information that underlies the changing patterns. This renders them
difficult to be used in astronomical applications that typically focus on
tracking changes in brightness in a single observing band, which are monitored
with sparse sampling, and in which the structural displacements between
individual image frames often exceed the dimensions of the instrumental point
spread function (PSF).

Astronomical images and high-resolution interferometric images   in particular
offer very limited (if any) opportunity to identify   ``ground control points''
or to build ``scene sets'', as employed   routinely in remote-sensing and
machine-vision applications \citep[{\em   cf.}][]{djamdji_geometrical_1993,
zheng_computational_1993,adams_dynamic_2003,zitova_image_2003,paulson_waveletbased_2010}. 
Structural patterns observed in astronomical images often do   not have a
defined or even preferred shape, which is an aspect   relied upon in a number of
the existing object recognition   algorithms \citep[{\em
e.g.},][]{agarwal+2003}. Astronomical objects normally do not   feature
sufficiently robust edges that would warrant applying  the edge-based detection
and classification commonly used in object-recognition methods
\citep{belongie_shape_2002}. In addition, astronomical images often feature
partially transparent optically thin structures in which multiple structural
patterns can overlap without full obscuration, which makes these   images even
more difficult to analyze using the algorithms   developed for the purposes of
remote-sensing and computer   vision. Because of these specifics, automated
analysis and tracking   of structural evolution in astronomical images remains
very   challenging, and it requires implementing a dedicated approach that can
address all of the main specific   characteristics of astronomical imaging of
evolving structures.

Currently, structural decomposition of astronomical images normally involves
simplified supervised techniques based on identification of specific features of
the structure \citep[{\em   e.g.}, ridge
lines,][]{hummel_c_1992,lobanov_a_1998,bach_u_2008}, analysis of image-
brightness profiles \citep[{\em   cf.}][]{lobanov_a_2001,lobanov_a_2003} , or
fitting the observed structure with a set of predefined templates ({\em e.g.},
two-dimensional Gaussian features). Two-dimensional cross-correlation has been
attempted only in very few cases \citep[{\em
e.g.},][]{biretta_j_1995,walker_vlba_2008}, each time requiring manual
segmentation of images, which imposed strong limitations on the number of
structural patterns that could be tracked.

In some particular situations, for instance, in images of extragalactic radio
jets, distinct structural patterns cover a variety of scales and shapes from
marginally resolved brightness enhancements caused by relativistic shocks
embedded in the flow \citep{zensus_j_1995,unwin_s_1997,lobanov_a_1999} to
thread-like patterns produced by plasma instability
\citep{lobanov_a_1998c,lobanov_a_2001,hardee_p_2005}. In the course of their
evolution, most of these patterns may rotate, expand, deform, or even break up
into independent substructures. This makes template fitting and correlation
analysis particularly challenging, and simultaneous information extraction on
multiple scales and flexible classification algorithms are required.

Deconvolution algorithms \citep[{\em cf.}][]{hogbom_j_1974,clark_b_1980}
extended to multiple scales \citep[{\em e.g.},][]{cornwell_t_2008} might in
principle be able to solve this task. However, comparing structures imaged at
different epochs is difficult as a result of the general non-uniqueness of the
solutions provided by deconvolution and because of an obvious need to group
parts of the solution together to describe structures that are substantially
larger than the image PSF.

A more robust approach to automatize identification and tracking of structural
patterns in astronomical images can be provided by a generic multiscale method
such as wavelet deconvolution or wavelet decomposition \citep[{\em
cf.}][]{starck_astronomical_2006}. While they are typically applied for image-
denoising and compactification, wavelets provide all ingredients necessary to
decompose the overall structure in an image into a robust set of statistically
significant structural patterns. This paper explores the wavelet approach and
presents a wavelet-based image segmentation and evaluation (WISE) method for
structure decomposition and tracking in astronomical images.  The method is
based on combining wavelet decomposition with watershed segmentation and
multiscale cross-correlation algorithms to treat temporal sparsity of
astronomical images, multiscale structural patterns, and their large
displacements between individual image frames.

The conceptual foundations of the method are outlined in Sect.~\ref{sc:concept}.
An algorithm for segmented wavelet decomposition (SWD) of structure into a set
of statistically significant structural patterns (SSP) is introduced in
Sect.~\ref{sc:swd}. A multiscale cross-correlation (MCC) algorithm for tracking
positional displacements of individual SSP is described in
Sect.~\ref{sect:structural_changes}. In Sect.~\ref{sect:tests}, WISE is tested
against simulated images of relativistic jets. In Sect.~\ref{sc:applications},
applications of WISE to astronomical images of parsec-scale radio jets in
3C\,273 and 3C\,120 are described and compared with results of conventional
structure analysis that was previously applied to these data. The results are
discussed and summarized in Sect.~\ref{sc:discussion}.

\section{Wavelet-based image structure evaluation (WISE) algorithm}
\label{sc:concept}

\subsection{Wavelet transform}

The wavelet transform is a time-frequency transformation that decomposes a
square-integrable function, $f(x)$, by means of a set of analyzing functions,
$\psi_{a, b}(x)$, obtained by shifts and dilations of a spatially localized
square-integrable wavelet function.

Different discrete realizations of the wavelet transform exist
\citep{mallat_a_1989,starck_astronomical_2006}. In the analysis presented here,
the {\em \`{a} trou} wavelet \citep{holschneider_a_1989, shensa_the_1992} was
used. This wavelet transform has the advantage of yielding a stationary,
isotropic, and shift-invariant transformation that is well-suited for
astronomical data analysis applications \citep{starck_astronomical_2006}.
Different scaling functions can be used with this transform
\citep{unser_splines_1999}. The choice of the scaling function is guided by the
specific properties of the image and the information required to be extracted
from the image \citep{ahuja_properties_2005}. In the following, we use the
B-spline scaling function (also called triangle function).

We treated digital astronomical images here as a sampled
representation ${f(x, y)}$ of the sky brightness distribution convolved with the
instrumental PSF. Following \cite{starck_astronomical_2006},
applying the \`{a} trou wavelet transform to ${f(x, y)}$ yields
\begin{equation} f(x, y) = \sum_{j=1}^{J} w_{j} (x, y) + c_{J}(x, y)\,.
\end{equation} Here, ${w_j(x, y)}$ is a set of resolution-related views of the
image, called {\em \textup{wavelet scales}}, which contain information on the wavelet
scale $j$ (corresponding to spatial scales from $2^{j-1}\,\omega$ to
$2^j\,\omega$, where $\omega$ is the limiting resolution in the image). The term
$c_J(x, y)$ is a smoothed array (a smoothed version of the original data
containing information of $f(x)$ on spatial scales $> 2^j$). The concept of the
spatial wavelet scale is therefore similar to the concept of a spatial
frequency, with smaller scales corresponding to higher frequencies and large
scales to lower frequencies.

\subsection{Conceptual structure of WISE}

To characterize the structure and structural evolution of an astronomical
object, the imaged object structure needs to be decomposed into a set of
significant structural patterns (SSP) that can be successfully tracked across a
sequence of images. This is typically done by fitting the structure with
predefined templates \citep[such as two-dimensional Gaussians, disks, rings, or
other shapes deemed suitable for representing particular structural patterns
expected to be present in the imaged region;][]{fomalont_e_1999,pearson_t_1999}
and allowing their parameters to vary. It is clear, however, that for a robust
structural decomposition made without a priori assumptions, the generic shape of
these patterns must be allowed to vary as well. To ensure this, a method is
needed that can automatically identify arbitrarily shaped statistically
significant structural patterns, quantify their significance, and provide robust
thresholding based on the significance of individual features.

The multiscale decomposition provided by the wavelet transform
\citep{mallat_a_1989} makes wavelets exceptionally well-suited to perform such a
decomposition, yielding an accurate assessment of the noise variation across the
image and warranting a robust representation of the characteristic structural
patterns of the image. To further increase the robustness of the method, the
multiscale approach is extended here to object detection, similarly to the
methodology developed for the multiscale vision model
\citep[MVM;][]{rue_a_1997,starck_astronomical_2006} in   related work on object
and structure detection \citep{shchikov_multiscale_2012,
seymour_multiresolution_2002}.  By combining these features, we have developed a
new, wavelet-based image structure evaluation (WISE) algorithm that is aimed
specifically at   the structural analysis of semi-transparent, optically thin
structures   in astronomical images. The method employs segmented wavelet
decomposition (SWD) of individual images into arbitrary two- dimensional SSP (or
image regions) and subsequent multiscale cross-correlation (MCC) of the
resulting sets of SSP. A detailed description of the method is given below.

\section{Segmented wavelet decomposition}
\label{sc:swd}

The segmented wavelet decomposition (SWD) comprises the following steps
to describe an image structure by a set of statistically
significant patterns:

\begin{itemize}
\item[1.]~A wavelet transform is performed on an image $I$  by
decomposing the image into a set of $J$ sub-bands (scales), $w_{j}$,
  and estimating the residual image noise (variable across the image).

\item[2.]~At each sub-band, statistically significant wavelet
  coefficients are extracted from the decomposition by thresholding
  them against the image noise.

\item[3.]~The significant coefficients are examined for local maxima,
  and a subset of the local maxima satisfying composite detection
  criteria is identified. This subset defines the locations of SSP in
  the image.

\item[4.]~Two-dimensional boundaries of the SSP are defined 
  by the watershed segmentation using the feature
  locations as initial markers.

\end{itemize} 

\noindent 
These steps essentially combine the MVM approach with watershed
segmentation and a two-level thresholding for the purpose of   yielding a robust
SSP identification procedure that would improve   the quality of subsequent
tracking of SSP that have been   cross-identified in a sequence of images of the
same object.

The SWD decomposition delivers a set of scale-dependent models (SDM), each
containing two-dimensional features identified at the respective scale of the
wavelet decomposition. The combination of all SDM provides a structure
representation that is sensitive to compact and marginally resolved features as
well as to structural patterns much larger than the FWHM of the instrumental PSF
in the image. Moreover, individual SSP identified at different wavelet scales
are partially independent, which allows for spatial overlaps between them and
can be used to improve the robustness and reliability of detecting structural
changes by cross-correlating multiple images of the same object.

\subsection{Determination of significant wavelet coefficients}
\label{sect:significant_coefs}

The statistical significance of a wavelet coefficient is given by its
probability to result from the noise in the image. This probability   is
determined using the multiresolution support technique
\citep{murtagh_image_1995}, which defines threshold $\tau_{j}$ above   which
wavelet coefficients are considered significant. The threshold   depends on the
noise characteristics in the image and on a false-discovery rate (FDR),
$\epsilon$. In the case of Gaussian noise, the   threshold can be defined as
$\tau_{j} = k_\mathrm{s} \sigma_{j}$,   where $k_\mathrm{s}$ is a factor that
depends on $\epsilon$.  The standard deviation, $\sigma_{j}$, of the noise at
wavelet scale $j$   is determined from the evaluation of the noise in the image
\citep{starck_astronomical_2006}. Choosing $k_\mathrm {s}=3$ gives   $\epsilon =
0.002$. Different techniques exist to handle other types   of noise, including,
for example, the use of the Anscombe transform   for Poisson noise. We refer to
\cite{starck_astronomical_2006} for a   complete review of noise treatment in
wavelet analysis of   astronomical images.

The application of the threshold condition yields a map of significant
coefficients for each wavelet scale:
\begin{equation}
m_{j}(x, y) = 
\begin{cases}
   w_{j}(x, y) &\mbox{ if $\abs{w_{j}(x, y)} >= k_\mathrm{s} \sigma_{j}$ }\\
   0 &\mbox{ otherwise }\,.
\end{cases}
\end{equation}

\subsection{Localization of significant structural patterns}
\label{sect:local_max}
A maximum filter is used to identify putative positions of SSP at each
scale of the wavelet decomposition. The filter comprises applying the
morphological operation of {\em \textup{dilation}} with a structuring element
of a desired size. The location of a local maxima occurs when the
output of this operation is equal to the original data value. This
defines a list of local maxima, $H_{j}$, at the scale $j$:
\begin{equation}
H_{j} = \{(x, y): \dilation(w_{j}(x, y)) = w_{j}(x, y)\}\,.
\end{equation}
The shape and size of the chosen structuring element affect the
smallest separation of two detected local maxima. For our specific application,
we use a diamond structuring element of a size that matched the scale at
which it is applied; with the minimum   size of two pixels.  Each of the lists
$H_j$ is clipped at a specific detection threshold, $\rho_j$.  This is done
recalling that for Gaussian noise, the detection level is proportional to
$\sigma_{j}$, hence $\rho_j = k_d \sigma_{j}$ can be set. For successful
detection thresholding, the condition $k_\mathrm{d}\ge k_\mathrm{s}$ must be
satisfied (with $k_\mathrm{d} = 4$--5 typically providing good thresholds).

The threshold clipping can be applied for defining $F_{j}$ as a group of
significant feature locations:
\begin{equation}
F_{j} = \{f = (x, y): (x, y) \in H \wedge \abs{w_{j}(x, y)} \geq k_\mathrm{d} \sigma_{j}\}\,,
\end{equation}
and these locations can be used for subsequently defining SSP
in the image.

\subsubsection{Identification of significant structural patterns}

\begin{figure}
  \centering
  \includegraphics{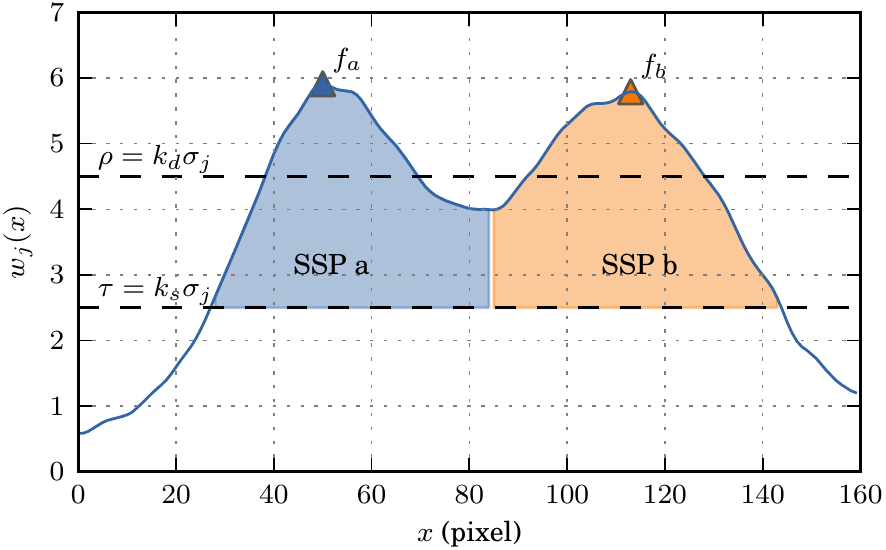}
  \caption{\label{fig:detection_principle}
    Schematic illustration of the method used for SSP localization, applied to 
    a one-dimensional case. The local maxima (triangle marker) are located using the maximum 
    filter and the SSP are associated with each of the local maxima by applying
    the watershed flooding algorithm. In this example the SSP ``a'' associated 
    with the position $f_a$ is defined by a region between 25 pix and 82 pix.}
\end{figure}

An SSP is defined as a 2D region of enhanced intensity extracted at a given
wavelet scale. To determine the extent and shape of individual SSP associated
with significant local maxima, image segmentation needs to be performed. The
segmentation relates each local maximum to a range of surrounding pixels that
can be considered part of this local intensity enhancement. The map of
significant coefficients $m_j$ is used for that purpose. The borders between
individual regions are determined from the common minima located between the
adjacent regions. This is achieved by watershed flooding
\citep{beucher_the_1993}\footnote{The watershed flooding earns its   name from
effectively corresponding to placing a ``water source'' in   each local minimum
and ``flooding'' the image relief from each of   these ``sources'' with the same
speed. The moment that the floods   filling two distinct catchment basins start
to merge, a dam is   erected to prevent mixing of the floods. The union of all
dams constitutes the watershed line.}. Figure~\ref{fig:detection_principle}
illustrates the application of the watershed segmentation in a one-dimensional
case.

The watershed segmentation is performed on $-m_{j}$ at all scales $j$
with $F_{j}$ as ``water sources'' , or markers. Each local maximum $f_a$ of
$F_{j}$ gives a region $s_{j, a}$ defined as
\begin{equation}
s_{j, a}(x, y) = 
\begin{cases}
   m_{j}(x, y) & \parbox[t]{.25\textwidth}{if $(x, y)$ is inside the watershed line of $f_a$ }\\
   0 & \parbox[t]{.25\textwidth}{ otherwise }\,.
\end{cases}
\end{equation}
The resulting SSP representation of an image at the
scale $j$ is finally derived as the group of regions:
\begin{equation}
 S_{j} = \{s_{j, a}: f_a \in F_{j}\}\,.
\end{equation}
An example of applying the SSP identification is shown in
Fig.~\ref{fig:jet_simultaion_wds} for a simulated
image of a compact radio jet.

\section{Multiscale cross-correlation}
\label{sect:structural_changes}

To detect structural differences between two images of an
astronomical object made at epochs $t_1$ and $t_2$, one needs to find
an optimal set of displacements of the original SSP (described by the
groups of SSP $S_{j,1}, j=1,...,J$) that would match the SSP in
the second image (described by $S_{j,2}$,
$j=1,...,J$). Cross-correlating $S_{j,1}$ and $S_{j,2}$ is a
natural tool for this purpose. There are two specific issues
that need to be addressed, however, to ensure that  the
cross-correlation analysis is reliable. First, a viable rule needs to be
introduced to identify the relevant image area across which the
cross-correlation is to be applied. The typical choices of using the
full image area or manually selecting the relevant fraction of the
image \citep[{\em cf.}][]{pushkarev_a_2012,fromm_c_2013} are not
satisfactory for this purpose. Second, the probability of false
matching needs to be minimized for features with sizes smaller than the
typical displacement between the two epochs.

These two requirements can be met by multiscale
  cross-correlation (MCC), which combines the structural and
  positional information contained in $S_j$ at all scales of the
  wavelet decomposition. The MCC uses a coarse-to-fine hierarchical
  strategy that is well known in image registration. This
  principle has first been used in \cite{vanderbrug_twostage_1977}
and \cite{witkin_signal_1987}, who used Gaussian pyramids. It
was then
  extended to the wavelet transform by
\cite{djamdji_geometrical_1993} and \cite{zheng_computational_1993}.
We refer to \cite{zitova_image_2003} and
\cite{paulson_waveletbased_2010} for a review on the different
  techniques developed in this area. However, none of theses
  algorithms can be directly applied for our purpose. The main reasons
  for this difficulty are the following:

\begin{enumerate}
  
  \item The images we consider are sparsely sampled (with structural displacements
   on the order of the PSF size or even larger) and do not offer a set of
   ``ground-control points'' that facilitate image registration (while this aspect
   is a critical feature of virtually all of the remote-sensing and computer-vision algorithms).

   \item The images are often dominated by optically thin structures (with the
   possibility of two or more independent structural features projected onto
   each other and often having different displacement or velocity vectors).

   \item The structural patterns do not have a defined or even preferred shape,
   and their shape may also vary from one image to another.
\end{enumerate}

\noindent
All these aspects call for a method that differs significantly
from the approaches used in the fields of remote sensing and computer vision.

\begin{figure}
  \centering
  \includegraphics{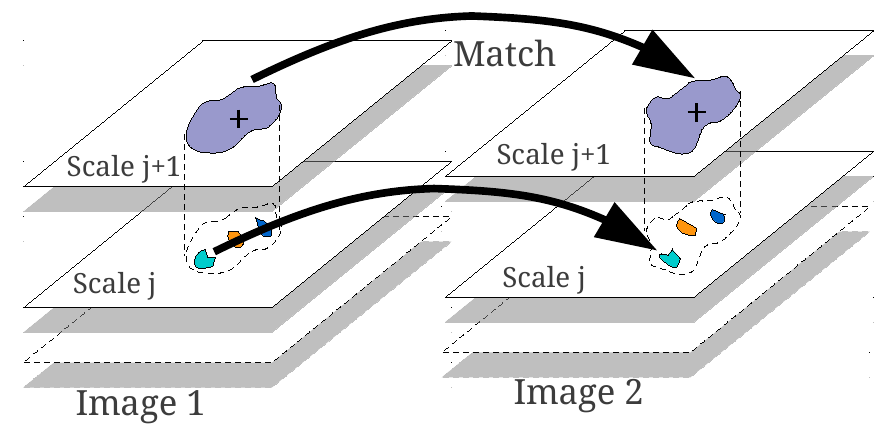}
  \caption{\label{fig:matching_principle} Illustration of the feature-matching method using a coarse-to-fine strategy. The
    calculated displacement at a higher (larger) scale is used to
    constrain the determination of the feature displacements at a
    lower (smaller) scale. In this particular example, the
      displacement of the SSP at scale $j+1$ is used as initial guess
      for displacements of its child SSPs at scale $j$. The initial
      guesses are subsequently refined by cross correlation.}
\end{figure}

Considering that SWD SSP at the wavelet scale $j$ have a typical size of
$2^{j}$, the largest displacement detectable on the scale $j$ must be
smaller than $2^{j}$. Identification of the structural displacements
can then begin from choosing $J$, the largest scale of the
wavelet decomposition, such that it exceeds the largest expected
displacement, but still satisfies the upper limit on $J$ given by the
largest scale containing statistically significant wavelet
coefficients. After correlating $S_{J,1}$ with $S_{J,2}$, the
respective correlations between $S_{j,1}$ and $S_{j,2}$ on smaller
scales are restricted to within the areas covered by $S_{J,1}$ and
$S_{J,2}$  in the two images. Alternatively, this approach
can also be used iteratively, restricting correlations on a given
scale $j$ to within the areas of the correlated features identified at
the $j+1$ scale. This algorithm is illustrated in
Fig.~\ref{fig:matching_principle}. Details of the procedure for
relating SSP identified at different scales are discussed
in the next section.

\subsection{Multiscale relations}
\label{sect:ms_relations}

Multiscale relations between SSP identified at different spatial
scales can be derived from the basic region properties. We note
again that the sizes of SSP identified at the scale $j$ are on the
order of $2^j$. Hence, any two individual SSP $s_a$ and $s_b$ of
$S_{j}$, identified around respective local maxima $f_a$ and $f_b$,
are separated from each other by at least $2^j$. This corresponds to
the inequality
\begin{equation}
\label{eq:ms_relation_features}
\norm{f_a - f_b} \gtrapprox 2^j, \forall f_a, f_b \in F_{j}, a \ne b\,. 
\end{equation}
If one determines a displacement $\Delta_{j + 1, b}$ of the SSP
$s_{j + 1, b}$ at the scale $j + 1$ between two epochs $t_1$ and
$t_2$, the following relation can be applied for the features of $F_{j}$
that are inside $s_{j + 1, b}$:
\begin{equation}
\label{eq:ms_relation_scales}
\Delta_{j, a} = \Delta^{j + 1}_{j, a} + \delta_{j, a}\,,
\end{equation}
for all $f_a \in F_{j}$ and $f_b \in  F_{j + 1}$, so that $s_{j + 1, b}(f_a) >
0$ and the condition $\Delta^{j + 1}_{j, a} = \Delta_{j + 1, b}$ is satisfied.
>From Eqs.~\eqref{eq:ms_relation_features} and~\eqref{eq:ms_relation_scales},
it also follows that
\begin{equation}
\label{eq:ms_relation_born}
\norm{\delta_{j, a}} < \frac{2^{j + 1}}{2}\,.
\end{equation}
Based on these relations, we adopted the following MCC algorithm to detect
structural changes between two images of an astronomical object:

\begin{enumerate}

\item The largest scale $J$ of a wavelet decomposition is chosen such
  that either the largest expected displacement is smaller than
  $2^J$ or $J$ corresponds to the largest scale with statistically
  significant wavelet coefficients.

\item Displacements of SSP features are determined at the largest
  scale $J$. For this calculation, all $\Delta^{J + 1}_{J, a}$ are
  set to zero, and $\Delta_{J, a} = \delta_{J, a}$ is calculated for
  each SSP.

\item At each subsequent scale $j$ ($j<J$), $\Delta^{j + 1}_{j, a}$
  is determined first by adopting the displacement $\Delta_{J, a}$
  measured at the $j + 1$ scale for the SSP in which the given
  $j$-scale region $s_{j, a}$ falls. Then the total displacement
  for this SSP is given by $\Delta_{j, a} = \Delta^{j + 1}_{j, a}
  + \delta_{j, a}$.
\end{enumerate}
In this algorithm, the only quantity that needs to be calculated
at each scale is the relative displacement $\delta_{j, a}$. This
quantity is bound by Eq.~\eqref{eq:ms_relation_born} and, 
within this bound, it can be determined reliably from the 
cross-correlation.

\subsection{Correlation criteria for MCC}

The correlation is calculated between a reference image $r$ and a
target image $t$, with the time order of the two images not playing any role. 
The correlation coefficients can be estimated using a number of
different correlation criteria (see \cite{giachetti_matching_2000} for a 
review). The most commonly used criteria are the cross correlation,
\begin{equation}
 C_\mathrm{CC}(r, t) = \sum r_i t_i\,,
\end{equation}
and the sum of squared differences,
\begin{equation}
 C_\mathrm{SSD}(r, t) = \sum (t_i - r_i)^2\,,\end{equation}
with i the pixel index.
The tolerance to an offset between the reference and the
target image is obtained by subtracting the mean value of the image
intensity (zero-mean correlation). Similarly, tolerance
to scale change is obtained by dividing the root-mean-square of the
image intensity (normalized correlation).

The MCC algorithm is required to be insensible to both the image offset and
scale change. The zero-mean normalized cross correlation (ZNCC) and zero-mean normalized sum of the squared difference (ZNSSD) can be applied for this
purpose. \cite{pan_equivalence_2010} have demonstrated that these two
criteria are equivalent. MCC uses the ZNCC method, based on its excellent
computational performance \citep{lewis_fast_1995}. The ZNCC is given by
\begin{equation}
C_\mathrm{ZNCC}(r, t) = \frac{\sum \overline{r_i} \overline{t_i}}{\sqrt{\sum \overline{r_i}^2 \sum \overline{t_i}^2}}\,,\end{equation}
with $\overline{r_i} = r_i - \overline{r}$, and $\overline{r}$
being the mean of $r$. This criterion reaches its highest unity value when the reference and target image are identical.

To detect structural changes between the reference and target
images, each single SSP $s_{j, a}$ of the reference image is cross
correlated with the target image. As every SSP is constrained to
be located within a specific region, one is actually only interested in
determining the correlation over that region. To achieve
this, a weighting function, $\omega$ is introduced, which is normalized
to unity and provides $\omega \equiv 0$ everywhere except inside the
region containing the SSP of interest. A weighted zero-mean
normalized cross correlation (WZNCC) can then be defined as
\begin{equation}
C_\mathrm{WZNCC}(r, t) = \frac{\sum \overline{r_i \omega_i} \overline{t_i \omega_i}}{\sqrt{\sum \overline{r_i \omega_i}^2 \sum \overline{t_i \omega_i}^2}}\,.
\end{equation}

\subsection{Detection of SSP displacements}
\label{sect:detect_displacements}

As shown in Sect.~\ref{sect:ms_relations}, the displacements of
individual features are determined starting from the largest scale and
progressing to the finest scale of the wavelet decomposition. For each
SSP at the scale $j$, an initial guess for its displacement is
provided by the displacement measured for the region at the scale
$j+1$, which includes the SSP in question. The initial guess is then
refined via the cross correlation.

\begin{figure}
  \centering
  \includegraphics{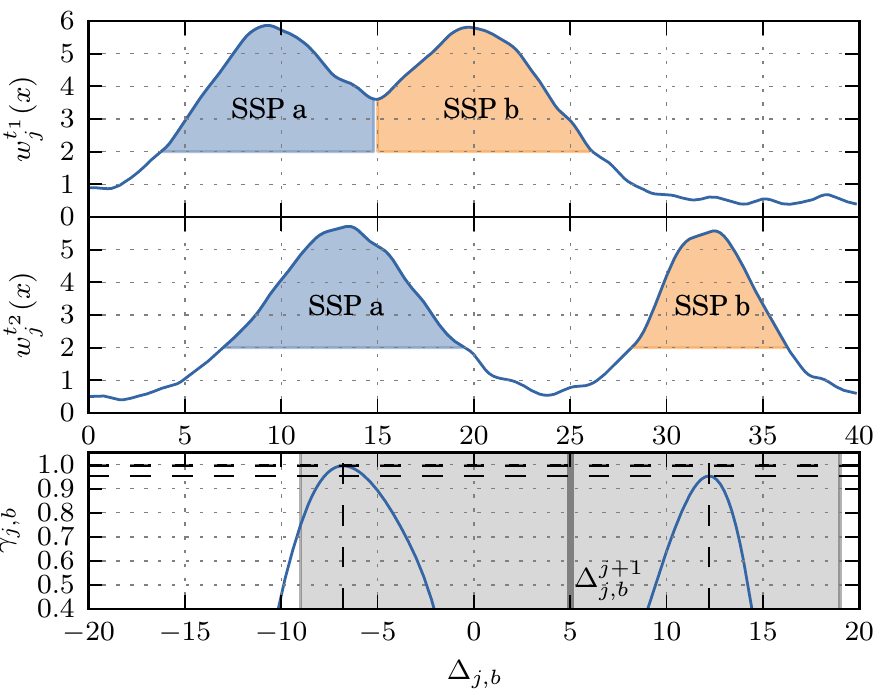}
  \caption{\label{fig:groupmatch_exemple} 
  Schematic illustration of the detection method used for displacement
  measurements in a one-dimensional case. In the upper two panels the wavelet
  decomposition at a scale $j$ of the reference (top panel) and target image is
  plotted, with two detected SSP marked with colors and letters. The
  x-axis of each panel is given in pixels. The result of WZNCC between SSP $b$
  and the target image is plotted below in the third panel. Two potential
  displacements are identified within the bounds (gray area) defined by
  Eq.~\eqref{eq:ms_relation_born} and the initial displacement guess $
  \Delta^{j+1}_{j, b}$ obtained from analysis at scale $j+1$.   To
  select the correct one and to reduce the chance for erroneous cross-correlation, the group motion of causally connected SSP (in this case SSP $a$
  and SSP $b$) is also included in the cross-correlation analysis, which results
  in the  identification of the displacement $\Delta_{j,b} = 12$.}
\end{figure}

This simple procedure is complicated by the fact that individual
SSP may merge, split, or overlap as a result of structural
changes occurring between the two observations. This means that the
displacement for which the cross correlation is maximized does not
necessarily provide the correct solution. Such a situation in
exemplified in Fig.~\ref{fig:groupmatch_exemple}. In this example,
SSP $b$ is moving faster than SSP $a$. As a
consequence, cross correlating the SSP $b$ at the epoch $t_1$
with $w_{j, t_2}$ yields the global maximum at $x^{t_2}_a$ and a local
maximum at $x^{t_2}_b$. The formal cross-correlation solution will be
incorrect in this case. To avoid such errors (or at least to
reduce their probability), it is necessary to cross-identify groups of
close SSP that can be related ({\em i.e.}, causally connected) to each
other in the two images. The cross correlation can then be applied to
these groups as well as to their individual members, so that a set of
possible solutions is found for all SSP, and the final solution is
determined through a minimization analysis applied to the entire group
of SSP.

At the first step of this procedure, subsets of features $G_{j}$ are
defined that are considered to be interrelated. As was discussed in
Sect.~\ref{sect:structural_changes}, at the scale $j$, $f_a$ is
independent from $f_b$ if $\norm{f_a - f_b} > 2^{j + 1}$. Then,
\begin{equation}
G_{j, u} = \lbrace x_i, x_i \in F_j \wedge \forall x_l \in F_j \setminus G_{j, u}, 
\norm{x_i - x_l} \geq 2^{j + 1} \rbrace\,,
\end{equation}
with
\begin{equation}
F_{j} = \sum_{u} G_{j, u}\,.
\end{equation}
At the second step, cross correlation is applied, yielding several
possible displacement vectors for each feature of such a
group. Considering the multiscale relations described in
Sect.~\ref{sect:ms_relations}, the correlation 
coefficients at $\delta = (\delta x, \delta y)$ can be calculated
for a given
feature $f_a$ of a group $G_{j, u}$:
\begin{equation}
\begin{split}  
\gamma_{j, a}(\delta x, \delta y) = C_\mathrm{WZNCC}(&s_{j, a}(x +\Delta x_{j, a}^{j + 1} + \delta x, y + \Delta y_{j, a}^{j + 1} + \delta y),\\
& w_j^{t_2}(x, y))\,,
\end{split}
\end{equation}
with $\norm{\delta} < 2^{j}$.

As illustrated by the example shown in
Fig.~\ref{fig:groupmatch_exemple}, for complex and strongly evolving
structures, it is possible that formally the best cross-correlation
solution provided by the largest $\gamma_{j,a,\mathrm{max}}$ may be
spurious. Hence, to avoid such spurious estimates of the
displacement vectors, all local maxima of
$\gamma_{j, a}$ that are above a certain threshold
$\kappa$ (with $\kappa$ usually set $\ge 0.8$) may be considered as possibly relevant solutions. 
These local maxima are found using the maximum filter
method described in Sect.~\ref{sect:local_max}.  

After identifying all relevant local maxima, the WZNCC of
the group of features is calculated for each possible group solution,
and we select the combination of individual displacement $\delta$
that maximizes the group correlation. This operation is repeated for
all groups of features $G_{j, u}$. This approach provides a robust
estimate of the statistically significant structural displacement
vectors across the entire image and at each structural scale.

In summary, our cross-correlation procedure comprised the following
main steps:
\begin{enumerate}

  \item Individual initial displacements and bounds are
  determined for each SSP using the relations
  of Eqs.~\eqref{eq:ms_relation_scales} and
\eqref{eq:ms_relation_born}.
  
  \item Groups of causally connected features are defined.

  \item Cross-correlation analysis is performed using the WZNCC
  for the groups and each of their elements, resulting in a set of
  potential displacements.

  \item The final SSP displacements are determined by selecting
  a combination of individual displacements that maximizes the overall
  group correlation.

\end{enumerate}

\subsection{Overlapping multiple displacement vectors}
\label{sect:max_displacement}

In images of optically thin structures, several physically
disconnected regions with different sizes and velocities may overlap,
causing additional difficulties for a reliable determination of
structural displacements (observations of transversely stratified jets
would be one particular example of such a situation).  Using the
partial independence of SWD SSP recovered at different wavelet scales,
the MCC method can partially recover these overlapping displacement
components. The largest detectable displacement inside a region is
determined by the largest wavelet scale $j$ for which this region can
be described by at least two SSP. Then, as described in
Sect.~\ref{sect:ms_relations}, the largest detectable displacement
would be $2^{j}$.  If velocity gradients or multiple velocity
components are expected inside this region, then this might not be
sufficient and the analysis might have to be started again at a wavelet scale
that describes the desired region by three or four different SSP.

The multiscale relations described in Sect.~\ref{sect:ms_relations}
rely on the assumption that SSP detected at a scale $j$ move, on
average, like their parent SSP detected at scale $j+1$. This
assumption sets limits for detecting different speeds at different
scales. Between two scales $j+1$ and $j$, this limit, determined
by Eq.~\eqref{eq:ms_relation_scales}, is on the order of $2^{j}$. As the
velocity difference approaches this limit, matching becomes more
difficult. If a very strong stratification or distinctly different
overlapping velocity components are expected, it is possible to relax
this constraint by introducing a tolerance factor $k_\mathrm{tol}$ in
Eq.~\eqref{eq:ms_relation_born},
\begin{equation}
\label{eq:ms_relation_born_with_tol}
\norm{\delta_{j, a}} < k_\mathrm{tol} * 2^{j}\,.
\end{equation}
This modification may increase the formal probability of spurious
matches, but the overall negative effect of introducing the tolerance
factor will be largely moderated by the cross-correlation part of the
algorithm. A similar limit applies if the gradient of velocity inside an
SSP is on the order of the SSP size.

\section{Testing the WISE algorithm}
\label{sect:tests}

To test the application of the WISE algorithm, simulated images of
optically thin relativistic jets were prepared that contained divergent
and overlapping velocity vectors manifested by structural
displacements generated for a range of spatial scales.

The simulated jet had an overall quasi-conical morphology, with a bright and
compact narrow end (``base'' of the jet) and smooth underlying flow pervaded by
regions of enhanced brightness (often called ``jet components'') moving with
velocities that varied in magnitude and direction. The underlying flow was
simulated by a Gaussian cylinder with FWHM $w_\mathrm{jet}$ evolving with the
following relation:
\begin{equation}
  w_\mathrm{jet}(z) = r_0 \frac{z}{z_0 + z} + r_1 \frac{z}{z_1} \tan(\phi_0)\,,
\end{equation}
where  $r_0$ is the width at the base of the jet, $z_0$ the axial $z$-coordinate 
of the jet base, and $z_1$ the $z$-coordinate of the point after which
$w(z)$ increase linearly with an opening angle of $\phi_0$,
and intensity $i_\mathrm{jet}$ evolving with the relation:
\begin{equation}
  i_\mathrm{jet}(z) = i_0 \left(\frac{z}{z_0}\right)^{\alpha}
,\end{equation}
where $\alpha$ is the damping factor.

\begin{figure}
  \centering
  \includegraphics{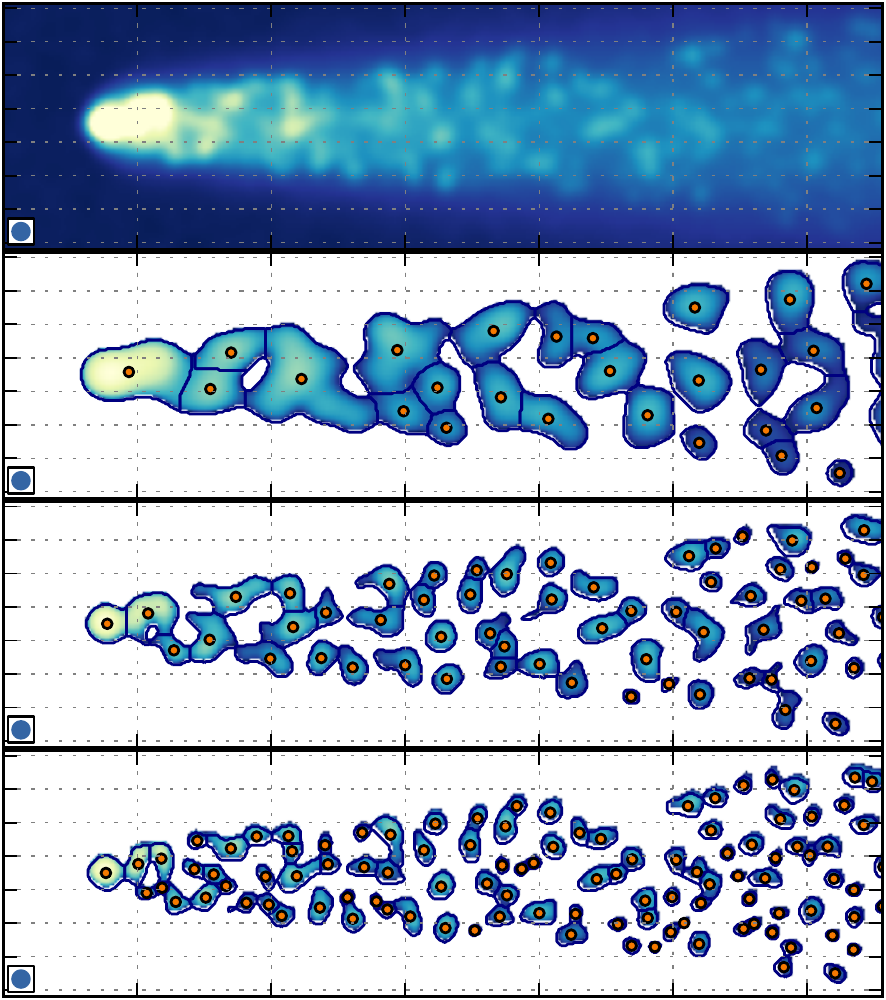}
  \caption{\label{fig:jet_simultaion_wds} Top panel shows a map of a
    simulated radio jet as described in
    Sect.~\ref{sect:tests}. The lower panels show the
    resulting SWD decomposition obtained at scales of 1.6, 0.8, and
    0.4 beam size.  }
\end{figure}

The jet base was modeled by a Gaussian component located on the jet
axis, at the position $z_0$. The moving features, also modeled by
Gaussian components (with randomly distributed parameters), were added
in the area defined by the jet after $z_1$.  The resulting image was
finally convolved with a circular or elliptical beam to
study the effect of different instrumental PSFs on the WISE
reconstruction of the simulated structural displacements.  An example
of a simulated image together with the SSP detected with the SWD at three
different scales is shown in Fig.~\ref{fig:jet_simultaion_wds}.


To evaluate the performance of WISE, two sets of tests were
performed.  The first set consisted of testing the SWD algorithm for
sensitivity to features at low SNR (\textup{\textup{{\em sensitivity test}}})
and for
distinguishing close and overlapping patterns (\textup{{\em separation test}}).
At the second stage of testing, the full WISE algorithm (combining the
SWD and the MCC parts) was applied to evaluate the sensitivity of the
method of detecting spatial displacements of individual patterns (\textup{{\em
  displacement test}}). In the following discussion, we define
  the SNR of a feature as its peak intensity over the noise level in
  the image.

\subsection{Sensitivity test}
\label{sect:sensibility_test}
This test was designed to represent as closely as possible the generic
use of the SWD algorithm for detecting and classifying structural
patterns in astronomical images.  The test was performed on a simulated
image of a jet, as illustrated in
Sect.~\ref{sect:tests}. For this particular simulation a
circular PSF with a FWHM of 10 pixels was applied. The morphology of the
underlying jet was given by the initial width $r_0 = 5\,\mathrm{FWHM}$
and an opening angle of 8$^\circ$.

Superimposed on the smooth underlying jet background, Gaussian
features with different sizes and intensities were then added. The
features were separated widely enough from each other to avoid
overlapping. The SWD method was applied to the simulated image, and the
SWD detections were then compared to the positions, sizes, and
intensities of the simulated features.  For the purpose of comparison,
we also performed a simple direct detection (DD), which consist of
detecting local maxima that are above a certain threshold directly
on the image. Similarly as for the SWD detection, the threshold for
the DD method was set to $k_\mathrm{d} \sigma_\mathrm{n}$, where
$k_\mathrm{d}$ is the detection coefficient as defined in
Sect.~\ref{sect:local_max}, and $\sigma_\mathrm{n}$ the standard
deviation of the noise in the image. When we determined whether a detected
feature corresponded to a simulated one, we used a tolerance of 0.2 FWHM of the
beam size on the position.

Table~\ref{tab:detection_limit} compares the performance of   the two methods.
The SWD method successfully recovered 95\% of   the extended features at
$\mathrm{SNR} \gtrsim 6$ with a low false-detection rate (FDR) for $k_d \ge
4$. This makes it a reliable tool   for detecting the statistically significant
structures in   astronomical images. In this particular test the SWD method
outperform the DD method by a factor of $\approx 4$.

\begin{table}[ht]
\label{tab:detection_limit}
\centering
\caption{
    Performance of the WDS detection compared to a direct detection (DD) in terms
    of SNR at which at 95\% of the simulated features are detected.
    Results are obtained for features with sizes of 0.2, 0.5, and 1 FWHM of the beam and
    for different detection threshold factors $k_d$. The bottom row of the table shows the mean false-detection rate (FDR) found in
    each test.}
\begin{tabular}{ccccc}
\toprule
\toprule
 Feature size& \multicolumn{4}{c}{SNR at 95\% detection rate} \\
\cmidrule(lr){2-5}
$\mathrm{[FWHM]}$ & \multicolumn{3}{c}{WDS detection}           & DD    \\
\cmidrule(lr){2-4}
& $k_\mathrm{d}=5$ & $k_\mathrm{d}=4$ & $k_\mathrm{d}=3$ & $k_\mathrm{d}=4$ \\
\midrule
0.2  & 5.1     & 4.4       & 3.4      & 27.6      \\
0.5  & 5.7     & 4.8       & 4.6      & 29.4      \\
1    & 8.0     & 6.6       & 5.9      & 36.0      \\
\midrule
FDR  & 0         & 0.01      & 0.1       & 0.03     \\
\bottomrule
\end{tabular}
\end{table}

\subsection{Separation tests}
\label{sect:separation_test}

The separation tests were designed to characterize the ability of the
SWD method to distinguish two close features. In this test, the images
structure comprised two Gaussian components of finite size that
were partially overlapping.  The two components were defined by their
respective SNR, $S_1$ and $S_2$ and FWHM, $w_1$ and $w_2$, and they
were separated by a distance $\Delta_\mathrm{s}$. For the purpose of
quantifying the test results, the fractional component separation
$r_\mathrm{s} = 2 \Delta_\mathrm{s}/(w_1 + w_2)$ was introduced. The
tests determined the smallest $r_\mathrm{s}$ for various combinations
of the component parameters at which the two features are detected. The
performance of the SWD algorithm was again compared with results from
applying the DD method introduced in
Sect.~\ref{sect:sensibility_test}.

\begin{figure}
  \centering
  \includegraphics{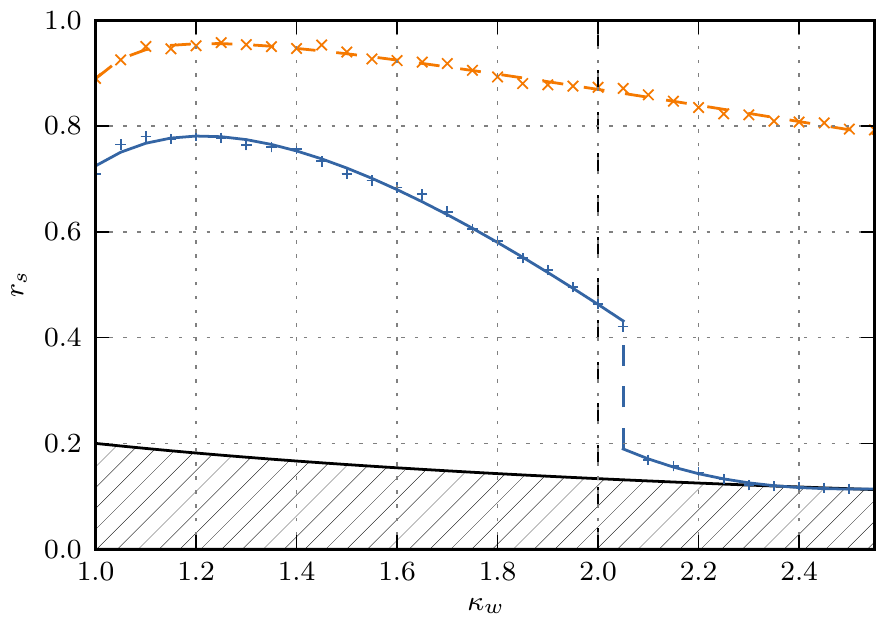}
  \caption{\label{fig:seperation_test_width}
  Characterization of the separability, $r_\mathrm{s}$, of two close features with
  varying FWHM ratio, $\kappa_\mathrm{w}$. Separation limit is determined for the SWD
  method (blue cross) and a direct detection method (yellow cross) as
  introduced in Sect.~\ref{sect:sensibility_test}. The gray hatched area
  is the region in the plot for which the separation between the two
  features is lower than 2 pixels.}
\end{figure}

In the first separation test, the ratio
$\kappa_\mathrm{w}=w_{1}/w_{2}$ was varied while setting $S_1 = S_2 =
 20$. Note that the features partially overlapped at their half-maximum
level for $r_\mathrm{s} \le
\kappa_\mathrm{w}/(1+\kappa_\mathrm{w})$. The results of this test are
shown in Fig.~\ref{fig:seperation_test_width}, with SWD always
performing better than the DD. In addition to this, the evolution of
smallest detectable $r_\mathrm{s}$ with $\kappa_\mathrm{w}$ indicates
two different regimes for SWD. For $1 \le \kappa_\mathrm{w} < 2$, SWD
progressively outperforms DD, with the difference between the two
increasing as $\kappa_\mathrm{w}$ increases. At $\kappa_\mathrm{w}
\ge 2$, SWD undergoes a fundamental transition, with both features
ultimately being always detected (at the 2-pixel separation limit).
This is the result of the multiscale capability of the SWD to identify and
separate power concentrated on physically different scales.

In the second separation test, the ratio $\epsilon_\mathrm{s} = S_{1}/S_{2}$ was
varied for features with $w_1 = w_2 = 10$\,pixels. The result demonstrates that
SWD performs better   than DD, with improvement factors rising from 30 \% to 50
\% with   increasing SNR ratio $\epsilon_\mathrm{s}$.

Both tests show that SWD is successful at resolving out two close and partially
overlapping features.  Assuming that the simulated component width $w_2$ in both
tests is similar to the instrumental PSF,  $r_\mathrm{s}$ can be interpreted as
$\approx 2/(1+\kappa_\mathrm{w})$\,PSF, implying that SWD successfully
distinguishes two marginally resolved features separated by 
$\approx 0.35\,\mathrm{PSF}\,(1+\epsilon_\mathrm{s})^{3/2}/(1+\epsilon_\mathrm {s}^2)^{1/2}$
, which is close to the expected limit
\[
r_\mathrm{s,lim} \approx \frac{2}{\sqrt{\pi}} \ln\left[\frac{S_2(1+\epsilon_\mathrm{s})+1}
{S_2(1+\epsilon_\mathrm{s})}\right]^{1/2} \frac{(1+\epsilon_\mathrm{s})^2}{\sqrt{1+\epsilon_\mathrm{s}^2}}\,\times\, \mathrm{PSF}
\]
for resolving two close features \citep[{\em cf.}][]{bertero_m_1997}.

\subsection{Structural displacement test}

These tests used the full WISE processing on a set of two simulated jet
images, first using the SWD algorithm to identify SSP features in each
of the images, then applying the MCC algorithm to cross-correlate
the individual SSP and to track their displacements from one image to
the other. The jet images were simulated using the procedure described
in the beginning of this section. A total of 500 elliptical features
were inserted randomly inside the underlying smooth jet, with their SNR
spread uniformly from 2 to 20 and the FWHM of the features ranging uniformly
from 0.2 to 1 beam size. The simulated structures were convolved with
a circular Gaussian (acting as an instrumental PSF) with a FWHM of 10
pixels.   A damping factor $\alpha$ of -0.3 was used.

Positional displacements were introduced to the simulated features in the second
image. The simulated displacements have both regular and stochastic (noise) components
introduced as follows:
\begin{equation}
\Delta_x = f_x(x) + G_x\,, \quad 
\Delta_y = f_y(x) + G_y\,,
\end{equation}
where $f_x$ and $f_y$ are the regular components of the displacement,
and $G_x$ and $G_y$ are two random variables following the Gaussian
distributions described by the respective means $<G_x>$, $<G_y>$ and
standard deviations $\sigma_x$, $\sigma_y$.
After the two images were generated, SSP were detected independently in each 
of them with the SWD and were subsequently cross-identified with the MCC.

\subsubsection{Accelerating outflow}

\begin{figure}
  \centering
  \includegraphics{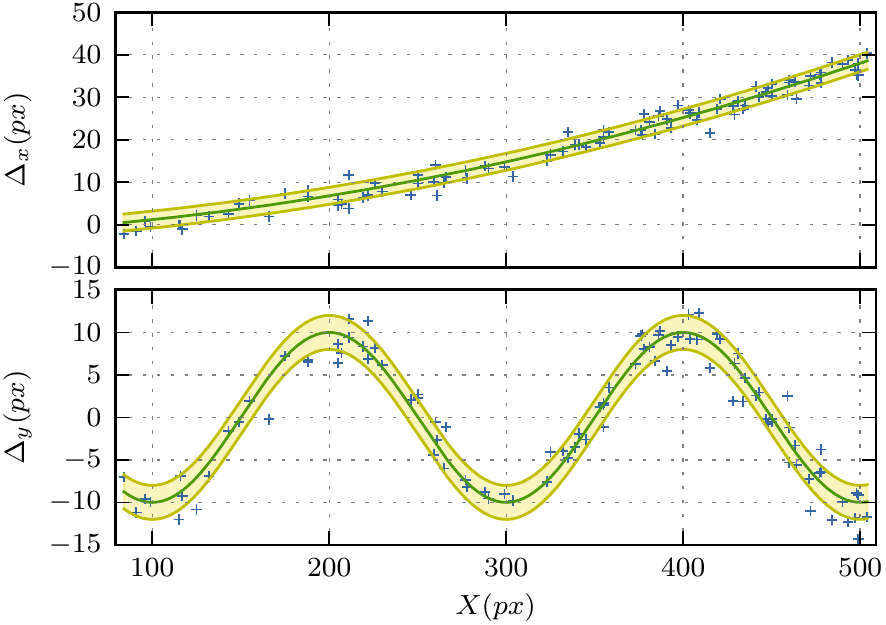}
  \caption{\label{fig:kinematic_simu_acc_big_sig2_full} WISE
    decomposition and analysis of a simulated jet with an accelerating
    sinusoidal velocity field. The input velocity field (green line)
    is defined analytically and modified with a Gaussian stochastic
    component with an r.m.s of 0.28\,FWHM of the convolving beam. The
    r.m.s. margins due to the stochastic component are represented by
    the yellow-shaded area. A total of 87\% of all detected SSP have been
    successfully matched by WISE. The detected positional changes
    (blue crosses) show r.m.s. deviations of 0.19 and 0.20 FWHM (in
    $x$ and $y$ coordinates, respectively) from the simulated
    sinusoidal field.}
\end{figure}

\begin{figure}
  \centering
  \includegraphics{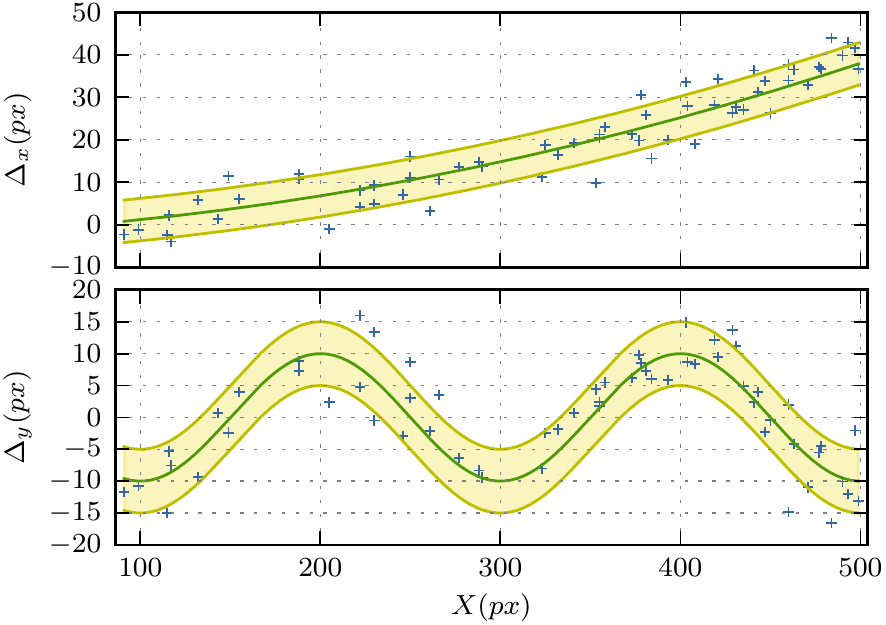}
  \caption{\label{fig:kinematic_simu_acc_big_sig5_full} Same as in
    Fig.~\ref{fig:kinematic_simu_acc_big_sig2_full}, but for the
    simulated stochastic component with an r.m.s. of 0.71\,FWHM of the convolving
    beam. The total of 54\% of all identified SSP have been
    successfully matched between the two simulated images.  The
    respective r.m.s of the deviations of the detected displacements
    from the analytic sinusoidal velocity field are 0.43\,FWHM and
    0.42\,FWHM, in $x$ and $y$ coordinates, respectively. }
\end{figure}

This displacement test explores a kinematic scenario describing an accelerating
axial outflow with a sinusoidal velocity component transverse to the main flow direction:
\begin{equation}
\label{eq:acc_sinus}
f_x(x) = a + b x + c x^2\,, \quad
f_y(x) = d \cos\left(\frac{2\pi\, x}{T}\right) \,.
\end{equation}
Results of the WISE application are shown in
Figs.~\ref{fig:kinematic_simu_acc_big_sig2_full}--\ref{fig:kinematic_simu_acc_big_sig5_full}
for $a=-2$, $ b=0.02$, $c=0.00012$, $d=10$, $T=200$, for the stochastic
displacement components with $\sigma_x = \sigma_y = 2$ and $\sigma_x = \sigma_y = 5$ (0.2\,FWHM and 0.5\,FWHM),
respectively (all linear quantities are expressed in pixels). The
largest expected displacement between the two images is 40 pixels. The
WISE analysis was performed on scales 2--6 (corresponding to 4--64 pixels).

The comparison between the simulated displacements and the displacements
detected by WISE  reveals an excellent performance of the matching algorithm. To
assess this performance, we computed the root mean square of the discrepancies
between the simulated and detected displacements:
\begin{equation}
e_x = \sqrt{\frac{1}{N}\sum_{i=1}^N{\left(\Delta x_{i} - f_x(x_i)\right)^2}}
\end{equation}
\begin{equation}
e_y = \sqrt{\frac{1}{N}\sum_{i=1}^N{\left(\Delta y_{i} -
f_y(x_i)\right)^2}}\,,
\end{equation}
where $\Delta x_i$, $\Delta y_i$ are the measured $x$ and $y$
components of the displacement identified for the $i^\mathrm{th}$
simulated component, and $x_i$ is the position of that component
along the $x$ axis in the first simulated image.  The $e_x$ and $e_y$
determined from the WISE decomposition do not exceed the $\sigma_x$
and $\sigma_y$ of the simulated data. For the first we obtain $e_x = 0.19$ and $e_y = 0.20$,
while for the second test, we obtain $e_x = 0.43$ and $e_y = 0.42$.
The number of positively matched features 
decreases with increasing stochastic component of the displacements, but the
errors of WISE decomposition always remain within the bounds determined
by the simulated noise.

These comparisons indicate that WISE performs very well even in the
case of relatively large spurious and random structural changes (which
may result from deconvolution errors, phase noise, and incompleteness
of the Fourier domain coverage by the data). Because such
spurious displacement is expected at a level of $\lesssim
\mathrm{FWHM}/\sqrt{\mathrm{SNR}}$, WISE should be able to reliably
identify displacement in regions detected at $\mathrm{SNR}\gtrsim 4$.

\subsubsection{Two-fluid outflow}

The purpose of this test was to investigate the possibility of   using WISE to
detect multiple velocity components in optically thin   materials in which
structures moving at different speeds overlap. To simulate such a two-fluid outflow, the initial set   of features was divided into two groups $F_1$
and $F_2$. The SNR and   FWHM of these features were derived from the same
distribution. The   simulated displacements, $\Delta_{1,2}$, were oriented
longitudinally   (along the $x$-axis) in the outflow and were the same in
all of   the features of a given group, with $\Delta x_1 = a$ and $\Delta x_2 =
b$, for the feature in $F_1$ and $F_2$, respectively.

\begin{figure}
  \centering
  \includegraphics{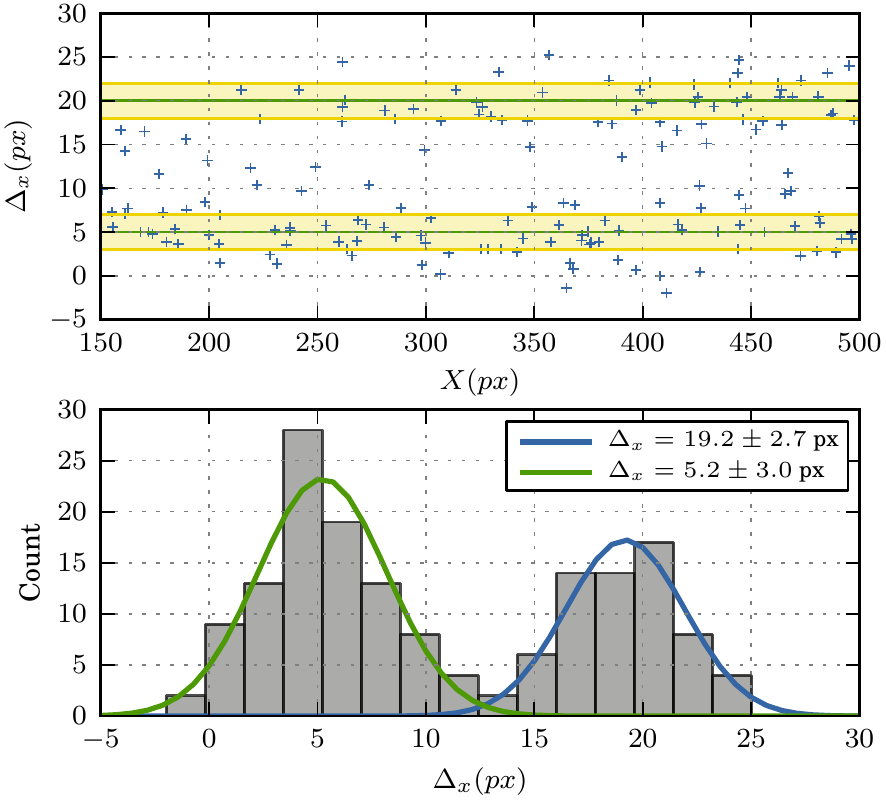}
  \caption{
  WISE decomposition and analysis of a simulated jet with two speed components.
  Two groups of features evolving at two different speeds are simulated. Upper
  panel: Blue crosses show the displacements detected by WISE from successive
  pairs in four simulated images (epochs). Green lines and orange shades
  indicate the two simulated displacements and r.m.s. of the Gaussian stochastic
  component. Lower panel: The histogram of the detected positional changes reveals
  two distinct components of the speed. The mean values of the speed and their
  r.m.s., indicated in the top right corner, agree well with the simulated
  displacements $\Delta x_1 = 5$\,px and $\Delta x_2 = 20$\,px.}

\label{fig:kinematic_simu_two_speed_sig2} 
\end{figure}

Results of the WISE application are shown in
Fig.~\ref{fig:kinematic_simu_two_speed_sig2} for $a=5$\,px, $b=20$\,px and
$\sigma_x = \sigma_y = 2$\,px (with a PSF size of 10\,px). It is expected that
a combination of more than two epochs is required to obtain
enough positional changes to distinguish the two different values of the speed.
In this case, combining the total of four epochs is found to be necessary. The
speeds of individual features were determined through a statistical analysis of
the detected displacements. The distribution of the detected displacements shown
in Fig.~\ref{fig:kinematic_simu_two_speed_sig2} is bimodal, with two clearly
separated peaks, and it can be used to derive the mean speed and its r.m.s.
for each of the two simulated flow components. This yields $\Delta x_1 = 19.2
\pm 2.7$\,px and $\Delta x_2 = 5.2 \pm 3.0$\,px, which agrees well with the
displacements used to simulate the two components of the flow.

\section{Applications to astronomical images}
\label{sc:applications}

We have tested the performance of WISE on astronomical images by
applying it to several image sequences obtained as part of the MOJAVE
long-term monitoring program of extragalactic jets with very long
baseline interferometry (VLBI) observations \citep[][and references
therein]{lister_mojave_2013}. The particular focus of the tests is on
two prominent radio jets in the quasar 3C\,273 and the Seyfert galaxy
3C\,120. These jets show a rich structure, with a number of enhanced
brightness regions inside a smooth and slowly expanding flow. This
richness of structure on the one hand has always been difficult to analyze by means of fitting it by two dimensional Gaussian features,
on the other hand, it has always suggested that the transversely
resolved flows may manifest a complex velocity field, with velocity
gradients along and across the main flow direction \citep[{\em
cf.}][]{lobanov_a_2001,hardee_p_2005}.

The MOJAVE observations, with their typical resolution of 0.5
milliarcsecond (mas), transversely resolve the jets in both
objects and, in addition, they also reveal apparent proper
motions of 3 mas/year in 3C\,120 \citep{lister_mojave_2013}, which
makes these two jets excellent targets for attempting to determine the
longitudinal and transverse velocity distribution.
 
The WISE analysis was applied to the self-calibrated hybrid images
provided at the data archive of the MOJAVE
survey\footnote{www.physics.purdue.edu/MOJAVE}. The results of WISE
algorithm were compared to the MOJAVE kinematic modeling of the jets
based on the Gaussian model fitting of the source structure
\citep[see][for a detailed description of the kinematic
modeling]{lister_mojave_2013}.

\subsection{Analysis of the images}

For each object, the MOJAVE VLBI images were first segmented
using the SWD algorithm, with each image analyzed independently. The
image noise was estimated by computing $\sigma_j$ at each
wavelet scale, as described in
Sect.~\ref{sect:significant_coefs}. Based on these estimates, a
$3\sigma_j$ thresholding was subsequently applied at each scale.
This procedure provides a better account for the scale dependence of
the noise in VLBI images \citep{lal_array_2010,lobanov_a_2012}, which is
expected to result from a number of factors including the coverage of
the Fourier domain and deconvolution.

Following the segmentation of individual images, MCC was
performed on each consecutive pair of images, providing the
displacement vectors for all SSP that were successfully cross-matched. The images were aligned at the position of the SSP that
was considered to be the jet ``core'' (which is typically, but not
always, the brightest region in the jet). This was done to
account for possible positional shifts resulting from self-calibration
of interferometric phases and for potential positional shifts (core
shift) due the opacity at the observed location of the jet base
\citep{lobanov_a_1998,kovalev_y_2008}.

For SSP that were cross-identified over a number of observing
epochs, the combination of these displacements provided a
two-dimensional track inside the jet. The track information from
several scales was also combined whenever a given SSP was
cross-identified over several spatial scales.

\begin{figure*}[ht]
  \centering
  \includegraphics{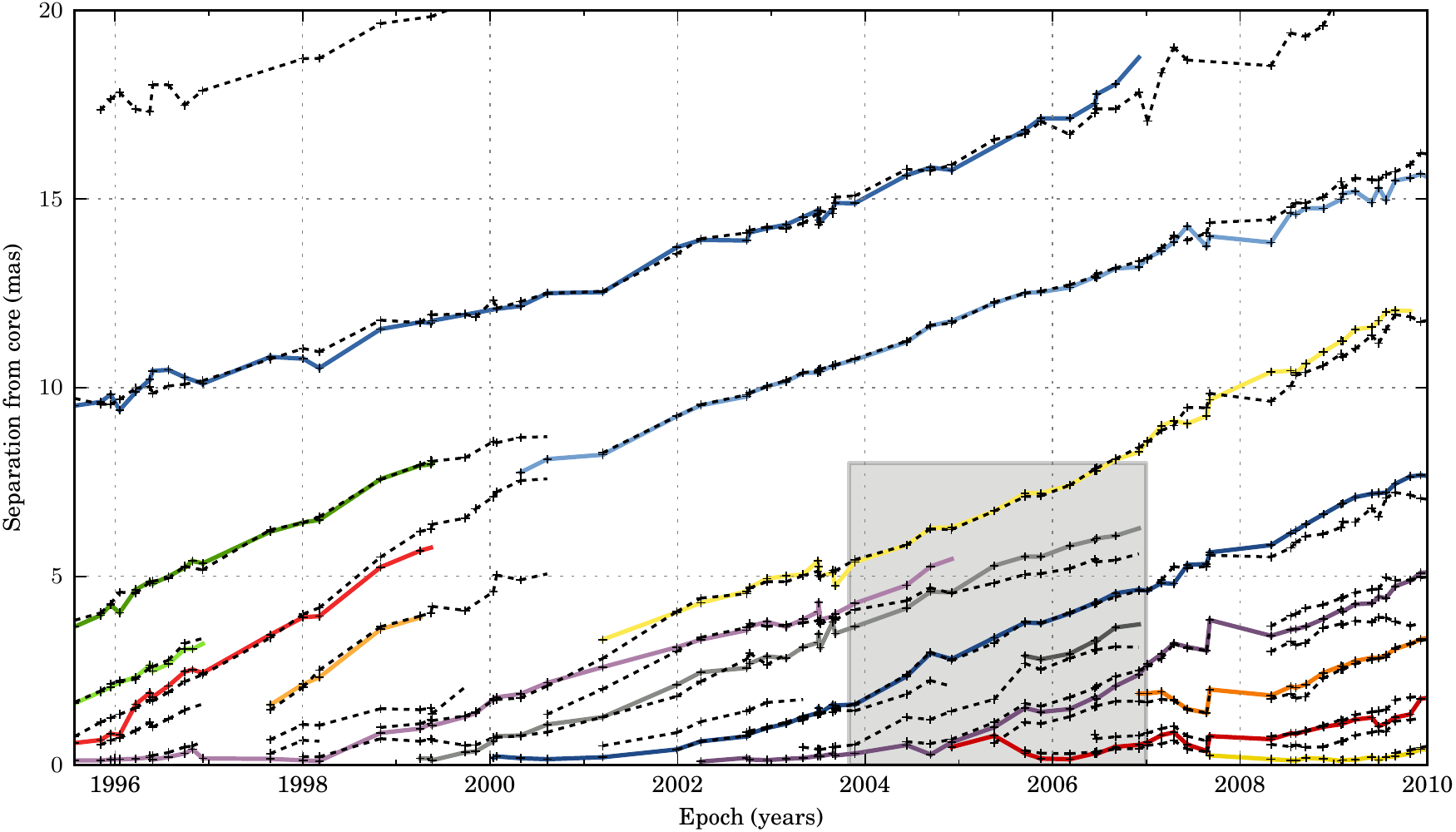}
  \caption{\label{fig:3c273_mojave_dfc_final} Core separation plot of
    the most prominent features in the jet of 3C\,273.  The model-fit
    based MOJAVE results (dashed lines) are compared with the WISE
    results (solid lines) obtained for SWD scales 3 and 4
    (selected to match the effective resolution of WISE to
    that of the Gaussian model fitting employed in the MOJAVE
    analysis). A detailed analysis, also including SWD scales 1
    and 2, was performed for the observations made between 11/2003
    and 12/2006 (gray box); the results are shown in
    Fig.~\ref{fig:3c273_mojave_dfc_200456}. }
\end{figure*}

\subsubsection{Jet kinematics in 3C\,273}
\label{sect:3c273}

The MOJAVE database contains 69 images of 3C\,273, with the
observations covering the time range from 1996 to 2010 and providing,
on average, one observation every three months. The SWD was
performed with four scales, ranging from 0.2 mas (scale 1) to
1.6 mas (scale 4).

For the MCC part of WISE, the individual images were aligned at the positions of
their respective strongest and most compact components (``core'' components) as
identified by the MOJAVE model fits.   The kinematic evolution of most of the
detected SSP is fully represented by the MCC results obtained for a single
selected SWD scale. However, long-lived features in the flow could eventually
expand so much that the wavelet power associated with a specific SSP would be
shifted to a larger scale, and the full evolution of such a feature was
described by a combination of MCC applications to two or more SWD scales.

The core separations of individual SSP obtained from WISE
decomposition are compared in Fig.~\ref{fig:3c273_mojave_dfc_final})
with the results from the MOJAVE kinematic analysis based on the
Gaussian model fitting of the jet structure. To provide this
comparison, the effective resolution of WISE must be reduced by
excluding the scales 1--2 from the consideration.
Comparison of the MOJAVE and WISE results in
Fig.~\ref{fig:3c273_mojave_dfc_final} indicates that WISE detects
consistently nearly all the components identified by the MOJAVE model
fitting analysis, with a very good agreement on their positional
locations and separation speeds.

\begin{figure*}[ht]
  \centering
  \includegraphics[width=0.9\textwidth]{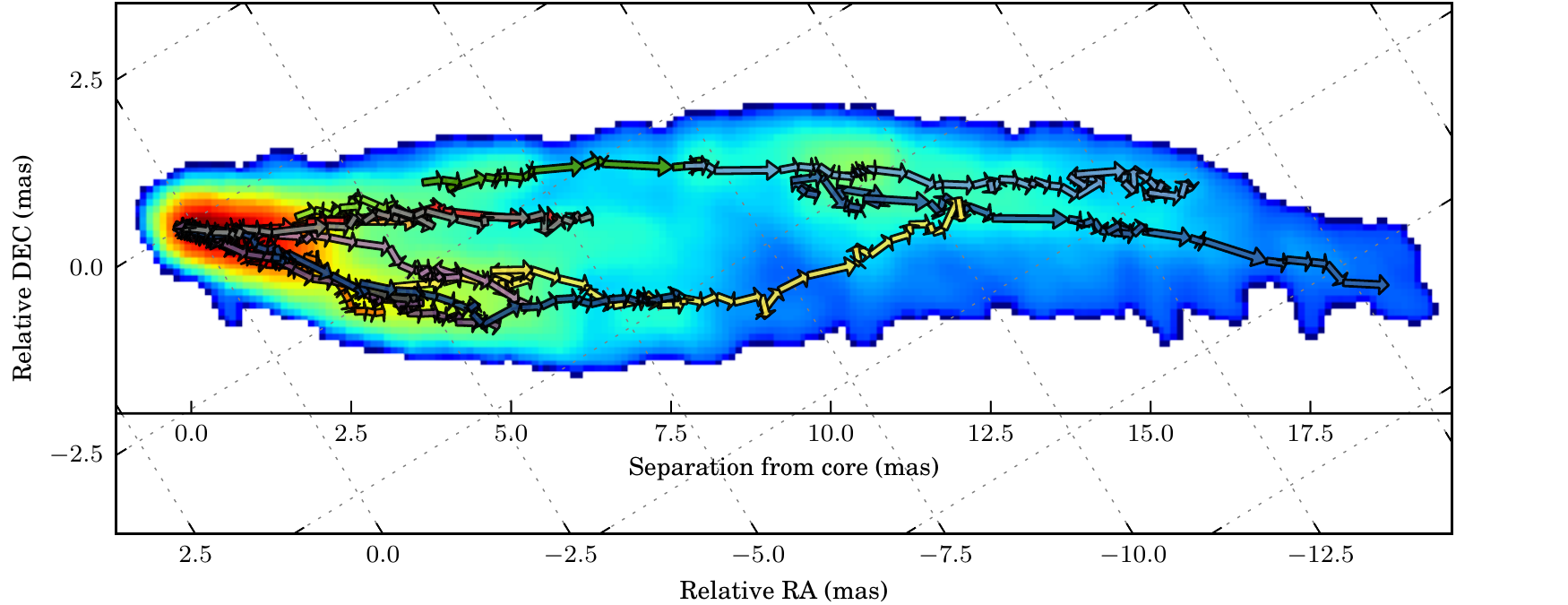}
  \caption{\label{fig:3c273_mojave_map_final} Two-dimensional tracks
    of the SSP detected by WISE at scales 3--4 of the SWD and compared
    in Fig.~\ref{fig:3c273_mojave_dfc_200456} with the features
    identified in the MOJAVE analysis of the images. The tracks are
    overplotted on a stacked-epoch image of the jet rotated by an angle of 
    0.55 radian. Colors distinguish individual SSP continuously tracked 
    over certain period of time. Several generic ``flow lines'' are clearly visible in the
    jet. These patterns are difficult to detect with the standard
    Gaussian model fitting analysis. The image is rotated. }
\end{figure*}

The two-dimensional tracks of the
WISE features detected with this procedure are shown in
Fig.~\ref{fig:3c273_mojave_map_final}, overplotted on a single-epoch image of
the jet. The displacement tracks clearly show several
``flow lines'' threading the jet, which can be associated with the
instability pattern identified in it \citep{lobanov_a_2001}. Some of
these tracks can also be identified in Gaussian model fitting, but
only if there is no substantial structural variations across the
jet. If this is not the case, Gaussian model fitting becomes too
expensive and too unreliable for the purpose of representing the
structure of a flow. In this situation, WISE provides a better way
to treat the structural complexity.  We therefore conclude
that WISE can be applied for the task of automated structural analysis
of VLBI images of jets (and similar sequences of images of objects
with evolving structure), yielding a great increase in the speed of
the analysis (analyzing 69 images of
3C\,273 took about ten minutes of computing, while the model fitting of
these images required several days of the researchers' time).

\begin{figure}[ht]
  \centering
  \includegraphics{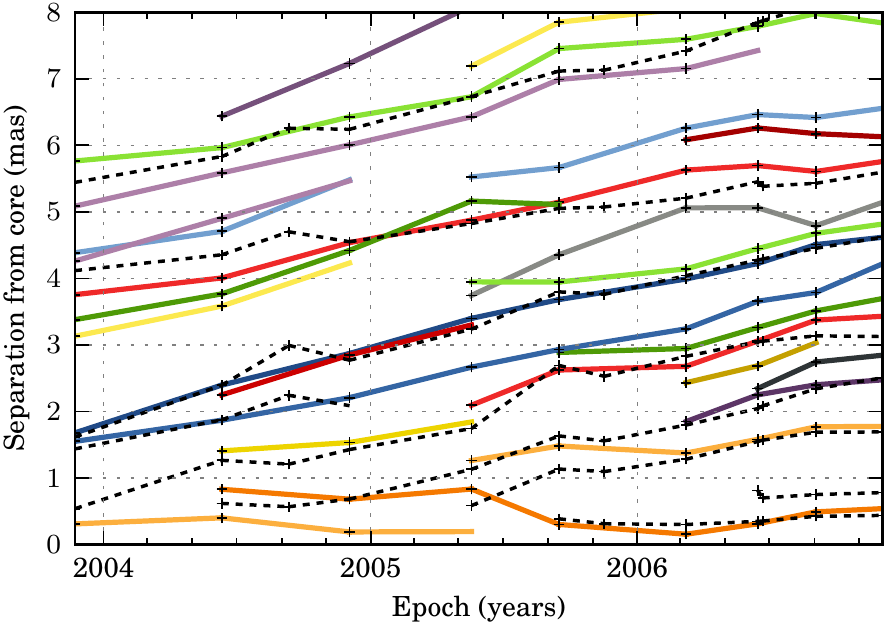}
  \caption{\label{fig:3c273_mojave_dfc_200456} Core separation plot of
    features detected in a detailed analysis of the jet of 3C\,273,
    which includes the SWD scales 1 and 2. Dashed lines show the
    MOJAVE model fit components, colored tracks present the SSP
    detected and tracked by WISE. At core separations $\gtrsim
    2$\,mas, WISE detects more significant features as
    the jet becomes progressively more resolved in the transverse
    direction (which also indicates that the structural description of the jet
    provided by Gaussian model fitting is no longer optimal).}
\end{figure}

\begin{figure*}[ht]
  \centering
  \includegraphics{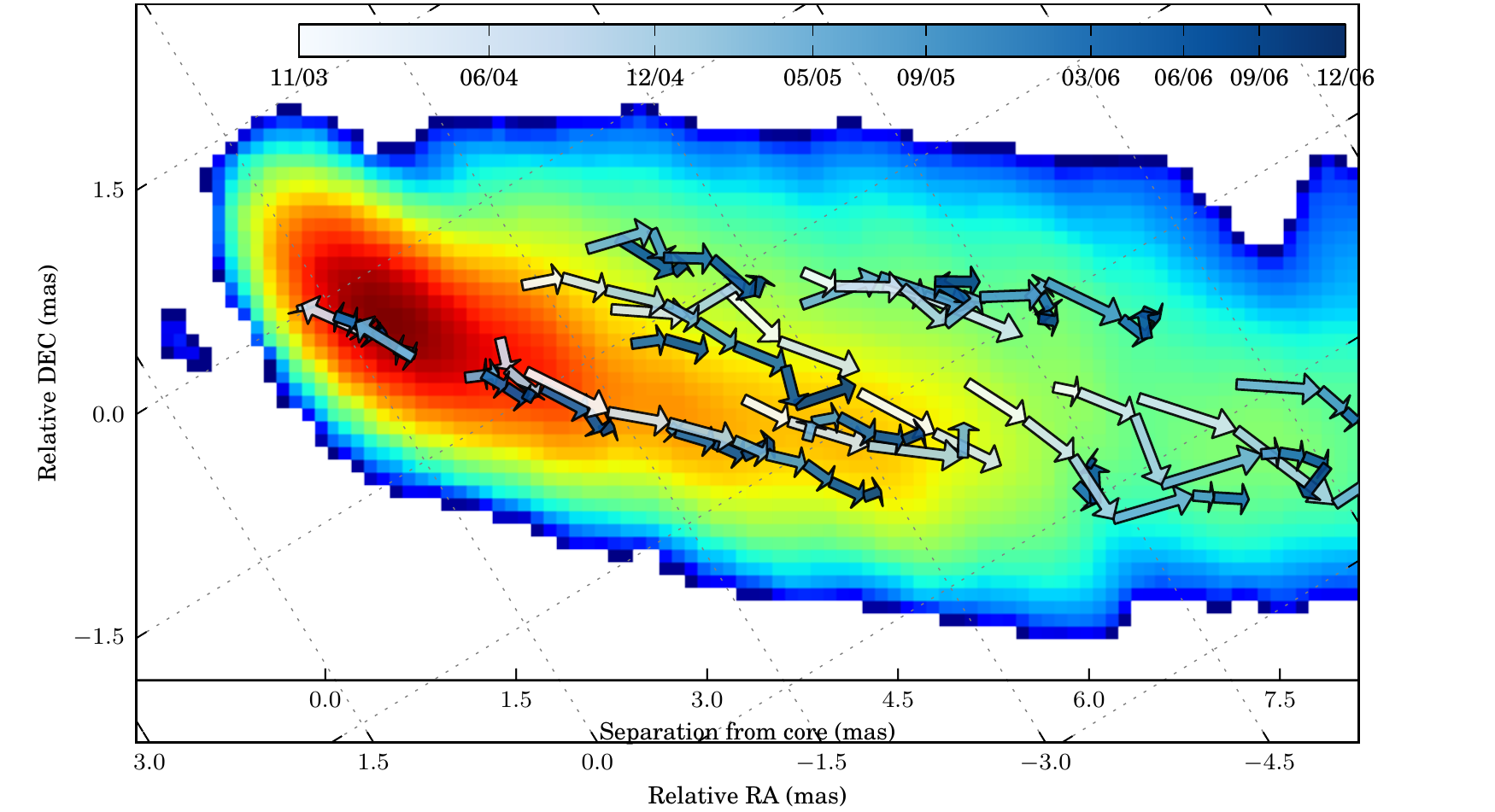}
  \caption{\label{fig:3c273_mojave_map_200456} 
  Two-dimensional tracks of SSP detected in 3C\,273 at the scale 2 of SWD for
  the epochs between 11/2003 and 12/2006. The tracks correspond to the features
  plotted in Fig.~\ref{fig:3c273_mojave_dfc_200456}. The colors of the
  displacement vectors indicate the measurement epoch as shown in the wedge at
  the top of the plot. The plot confirms the significant transverse structure in
  the jet, with up to three distinct flow lines showing strong and correlated
  evolution. The apparent inward motion detected in a nuclear region (0--
  0.3\,mas) is most likely an artifact of a flare in the jet core.}
\end{figure*}

However, WISE can certainly go beyond the resolution of Gaussian model
fitting by also including scales that are smaller than the
transverse dimension of the flow. An example of such an improvement is
shown in Fig.~\ref{fig:3c273_mojave_dfc_200456}, which focuses on
MOJAVE observations of 3C\,273 made between November 2003 and December
2006. At core separations larger than about 2 mas, WISE persistently
detects several features at locations where the Gaussian model fits
have been restricted to representing the structure with a single
component. This is a clear sign of transverse structure in the flow,
which is illustrated well by the respective displacement tracks shown
in Fig.~\ref{fig:3c273_mojave_map_200456}. These tracks provide strong
evidence for a remarkable transverse structure of the flow, with three
distinct flow lines clearly present inside the jet. These flow lines
evolve in a regular fashion, suggesting a pattern that may rise as a
result of Kelvin-Helmholtz instability, possibly due to one of the
body modes that have been previously identified in the jet based on a
morphological analysis of the transverse structure
\citep{lobanov_a_2001}. That analysis also implied that the flow
pattern probably rotates counterclockwise, and this rotation is consistent
with the general southward bending of the displacement vectors
(particularly visible in Fig.~\ref{fig:3c273_mojave_map_200456} at
distances of 4.5--6\,mas).

\subsubsection{3C120}
\label{sect:3c120}

The MOJAVE database for 3C\,120 comprises 87 images from observations
made in 1996--2010, averaging to one observation every three months (but
with individual gaps as long as one year). We prepared these images
for WISE analysis using the same approach as applied for
3C\,273. To ensure sensitivity to the expected displacements
of $\lesssim 3$\,mas between subsequent images, we applied SWD
on five scales, from 0.2 mas (scale 1) to 3.2 mas 
(scale 5).

\begin{figure*}
  \centering
  \includegraphics{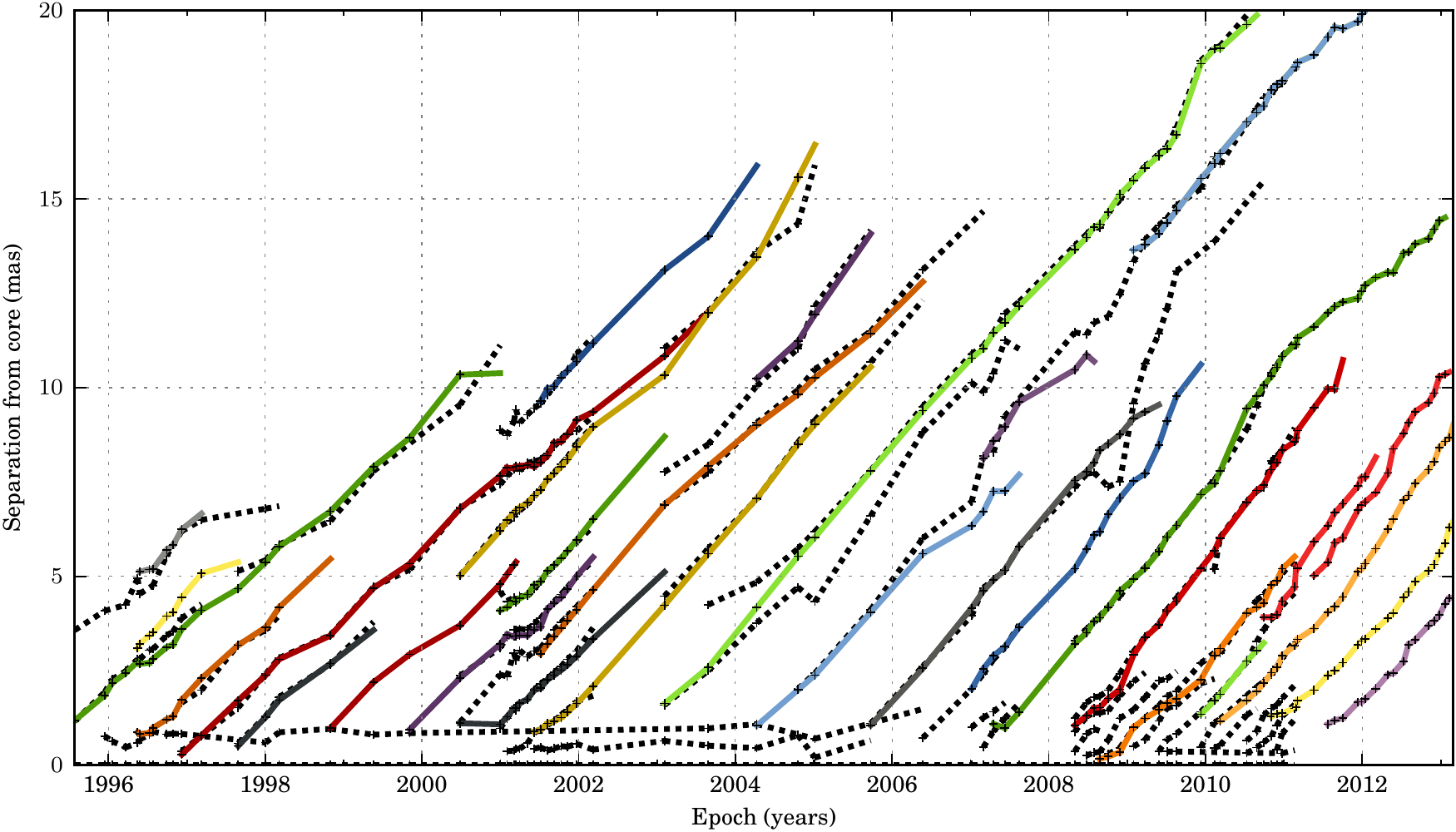}
  \caption{\label{fig:3c120_mojave_dfc_final}
    Core-separation plot of the features identified in the jet of 3C\,120.  
    The model-fit based MOJAVE results (dashed lines) are compared with the WISE
    results (solid lines) obtained for SWD scales 2 and 3
    (selected to match the effective resolution of WISE to
    that of the Gaussian model fitting employed in the MOJAVE
    analysis).}
\end{figure*}

Applied to the MOJAVE images of 3C\,120, WISE detects a total of 30
moving SSP. The evolution of 24 SSP is fully traced at the SWD scale 2
(0.4\,FWHM), and combining two SWD scales is required to describe the
evolution of the six remaining SSP.  The resulting core separations of
the SSP plotted in Fig.~\ref{fig:3c120_mojave_dfc_final} generally
agree very well with the separations of the jet components
identified in the MOJAVE Gaussian model fit analysis.  For the moving
features, displacements as large as $\sim 3$\,mas were reliably
identified during the periods with the least frequent observations.

The only obvious discrepancy between the two methods are the quasi-stationary
features that are identified in the MOJAVE analysis, but are absent from the
WISE results.  A closer inspection of the wavelet coefficients recovered at the
SWD scale 1 does not yield a statistically significant detection of an SSP at
the location of the MOJAVE stationary component either.

The stationary
feature identified in the MOJAVE analysis is often separated by less
than 1\,FWHM from the bright core, while it is substantially (factors
of $\sim 50$--100) weaker than the core.  This
extreme flux density ratio between two clearly overlapping components
may impede identifying the weaker feature against
the formal thresholding criteria of WISE. The fact that the Gaussian
model fitting was performed in the Fourier domain (not affected
by convolution) may have given it an advantage in this
particular setting. Subjective decision making during the model
fitting may also have played a role in the resulting structural
decomposition. 

Reaching a firm conclusion on this matter would require
assessing the statistical significance of the model fit
components identified with the stationary features and performing
the SWD separation test for extreme SNR ratios. We defer this to
future analysis of the data on 3C\,120, while noting again that WISE
has achieved its basic goal of providing an effective automated
measure of kinematics in a jet with remarkably rapid structural
changes.

\begin{figure*}
  \centering
  \includegraphics{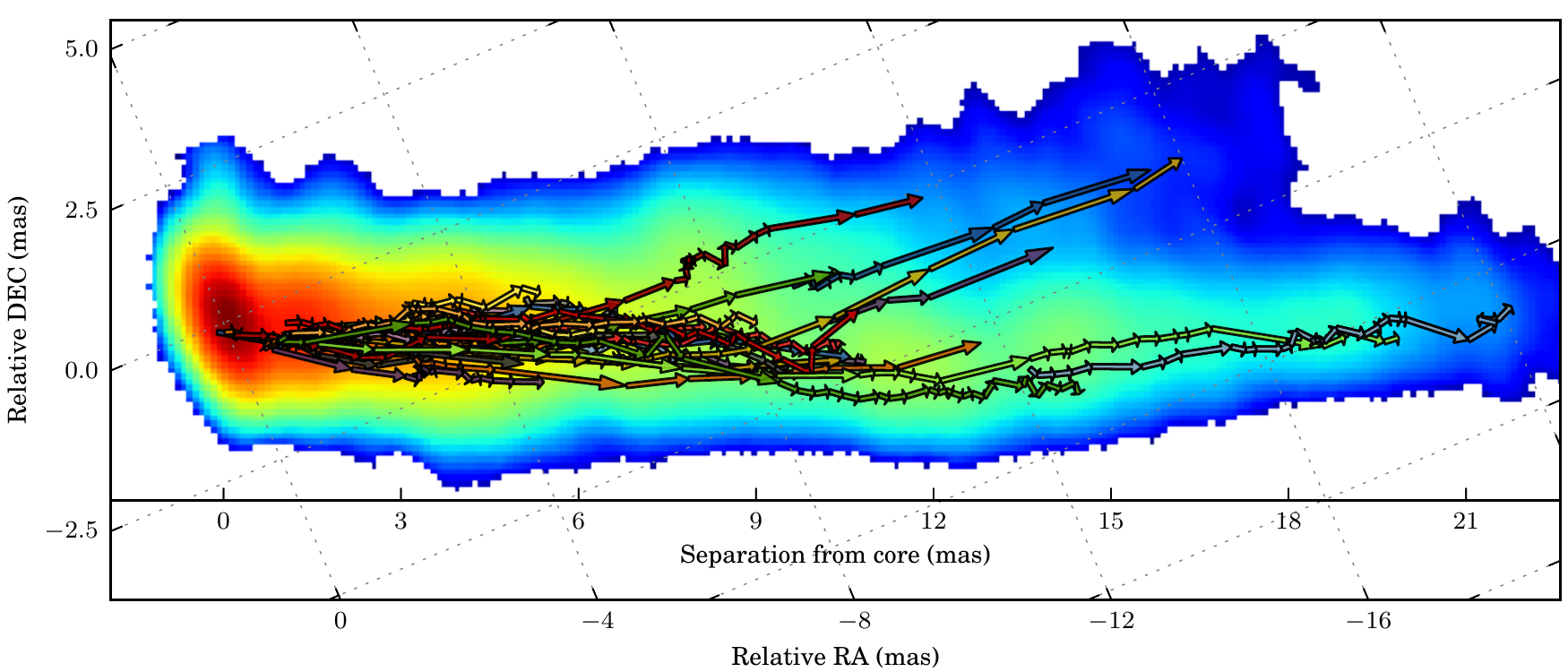}
  \caption{\label{fig:3c120_mojave_dfc_map} Two-dimensional tracks of
    SSP detected in 3C\,120 at scale 2 of SWD. The colored tracks
    correspond to the features plotted in
    Fig.~\ref{fig:3c120_mojave_dfc_final} in the same color. The tracks are
    overplotted on a stacked-epoch image of the jet rotated by an angle of 
    0.4 radian. The plot confirms the significant and evolving transverse
    structure in the jet, with individual tracks underlying the
    long-term evolution of the flow, which becomes particularly
    prominent at core separations of $\gtrsim 6$\,mas.}
\end{figure*}

The magnitude of the structural variability of the jet in 3C\,120 is
further emphasized in Fig.~\ref{fig:3c120_mojave_dfc_map}, which shows the
two-dimensional tracks of the SSP identified with WISE. The shape of
individual tracks suggests a helical morphology, consistent with the
patterns predicted from modeling the jet in 3C\,120 with
linearly growing Kelvin-Helmholtz instability \citep{hardee_p_2005}.
In this framework, the observed evolution of the component tracks is
consistent with the pattern motion of the helical surface mode of the
instability identified by \cite{hardee_p_2005} to have a wavelength of
$\sim 3.1$ jet radii and propagating at an apparent speed of $\sim
0.8\,c$. 

\cite{hardee_p_2005} also suggested that the structure of the flow is
strongly dominated by the helical surface mode, which may explain the
apparent lack of structural detail uncovered by WISE on the finest
wavelet scale. In this case, observations at a higher dynamic range
would be needed to reveal higher (and weaker) modes of
the instability developing in the jet on these spatial scales.
Altogether, the example of 3C\,120 again demonstrates the reliability
of the WISE decomposition and analysis of a structural
evolution that can be inferred from comparing multiple images of
an astronomical object.

\section{Conclusions}
\label{sc:discussion}

The WISE method we presented here offers an effective and
objective way to classify structural patterns in images of
astronomical objects and track their evolution traced by multiple
observations of the same object. The method combines automatic
segmented wavelet decomposition with a multiscale cross-correlation
algorithm, which enables reliable identification and tracking of
statistically significant structural patterns.

Tests of WISE performed on simulated images demonstrated its
capabilities for a robust decomposition and tracking of
two-dimensional structures in astronomical images.
Applications of WISE on the VLBI images of two prominent
extragalactic jets showed the robustness and fidelity of
results obtained from WISE compared with those coming from the ``standard''
procedure of using multiple Gaussian components to represent the
observed structure. The inherent multiscale nature of WISE allows it
also to go beyond the effective resolution of the Gaussian
representation and to probe the two-dimensional distribution of
structural displacements (hence probing the two-dimensional kinematic
properties of the target object).

In addition to this, the multiscale approach of WISE has several
other specific advantages. First, it allows simultaneous
detection of unresolved and marginally resolved features as well as
extended structural patterns at low SNR. Second, the method provides
a dynamic and structural scale-dependent account of the image noise
and uses it as an effective thresholding condition for assessing the
statistical significance of individual structural patterns. Third,
multiple velocity components can also be distinguished by the method
if these components act on different spatial scales -- this can be
a very important feature to study the dynamics of optically thin
emitting regions such as stratified relativistic flows,
with a combination of pattern and flow speed and strong transverse
velocity gradients.

Combining several scales also improved the cross-correlation
employed by WISE, ensuring a reliable performance of the method in the case
of severely undersampled data (with the structural displacement between
successive epochs becoming larger than the dimensions of the
instrumental point spread function).

In its present realization, WISE performs well on structures with
moderate extent, while it may face difficulties in correctly identifying
continuous structural details in which one of the dimensions is
substantially smaller than the other ({\em e.g.}, filamentary
structure and thread-like features). If the ratio between the largest
and smallest dimensions of this structure is lower than the ratio of
the largest and smallest scales of WISE decomposition, the continuity of
this structure may in principle be recognized. For more extreme cases,
WISE will break the structure into two or more SSP that are
considered independent. A remedy for this deficiency may be found in
considering groups of SSP during the MCC part of WISE, or by applying
more generic approaches to feature identification \citep[{\em e.g.},
shapelets; {\em cf.}][]{starck_astronomical_2006}.

Another probable, requiring additional attention is the scale crossing of
individual features that may occur as a result of expansion (as was illustrated
by the example of 3C\,273) or the particular evolution of a complex three-
dimensional emitting region projected onto the two-dimensional picture plane. At
the moment, this problem has to be treated manually and outside of WISE, but an
automated approach is clearly desired. One possibility here is to use the
wavelet amplitudes associated with the same SSP at different scales and to
select the dominant scale adaptively based on the comparison of these amplitudes
and their changes from one observing epoch to another.

Implementing this step may also require implementing a reliable error
estimation for the locations, flux densities, and dimensions of SSP
identified by WISE. This can be done on the basis of SNR estimates
performed at each individual scale of WISE decomposition. Generically,
it is expected that and SSP detected with a given SNR at a particular
wavelet scale $l_\mathrm{w}$ would have its positional and flux errors
$\propto l_\mathrm{w}/\mathrm{SNR}^{-1}$, while the error on the SSP
dimension would be $\propto l_\mathrm{w}/\mathrm{SNR}^{-1/2}$
\citep[{\em cf.}][]{fomalont_e_1999}. Such estimates can be
implemented as a zeroth-order approach, but a more detailed
investigation of the error estimates for the segmented wavelet
decomposition is clearly needed.

\begin{acknowledgements}
This research has made use of data from the MOJAVE database that is 
maintained by the MOJAVE team \citep{lister_m_2009}.
\end{acknowledgements}

\bibliographystyle{aa}
\bibliography{biblio}

\begin{thebibliography}{54}
\expandafter\ifx\csname natexlab\endcsname\relax\def\natexlab#1{#1}\fi

\bibitem[{Adams \& Williams(2003)}]{adams_dynamic_2003}
Adams, N.~J. \& Williams, C. K.~I. 2003, Image and Vision Computing, 21, 865

\bibitem[{Agarwal {et~al.}(2003)Agarwal, Awan, \& Roth}]{agarwal+2003}
Agarwal, S., Awan, A., \& Roth, D. 2003, IEEE Transactions on Pattern Analysis
  and Machine Intelligence, 26, 1475

\bibitem[{Ahuja {et~al.}(2005)Ahuja, Lertrattanapanich, \&
  Bose}]{ahuja_properties_2005}
Ahuja, N., Lertrattanapanich, S., \& Bose, N. 2005, Vision, Image and Signal
  Processing, IEE Proceedings -, 152, 659

\bibitem[{{Arulampalam} {et~al.}(2002){Arulampalam}, {Maskell}, \&
  {Gordon}}]{arulampalam_m_2002}
{Arulampalam}, M.~S., {Maskell}, S., \& {Gordon}, N.~and{Clapp}, T. 2002, IEEE
  Transactions on Signal Processing, 50, 174

\bibitem[{{Bach} {et~al.}(2008){Bach}, {Krichbaum}, {Middelberg}, \&
  {Alef}}]{bach_u_2008}
{Bach}, U., {Krichbaum}, T.~P., {Middelberg}, E., \& {Alef}, W.~and{Zensus},
  A.~J. 2008, in The role of VLBI in the Golden Age for Radio Astronomy

\bibitem[{Belongie {et~al.}(2002)Belongie, Malik, \&
  Puzicha}]{belongie_shape_2002}
Belongie, S., Malik, J., \& Puzicha, J. 2002, IEEE Transactions on Pattern
  Analysis and Machine Intelligence, 24, 509

\bibitem[{{Bertero} {et~al.}(1997){Bertero}, {Boccacci}, \&
  {Piana}}]{bertero_m_1997}
{Bertero}, M., {Boccacci}, P., \& {Piana}, M. 1997, in Lecture Notes in
  Physics, Berlin Springer Verlag, Vol. 486, Lecture Notes in Physics, Berlin
  Springer Verlag, ed. G.~{Chavent} \& P.~C. {Sabatier}, 1

\bibitem[{Beucher \& Meyer(1993)}]{beucher_the_1993}
Beucher, S. \& Meyer, F. 1993, Optical Engineering, 34, 433

\bibitem[{{Biretta} {et~al.}(1995){Biretta}, {Zhou}, \&
  {Owen}}]{biretta_j_1995}
{Biretta}, J.~A., {Zhou}, F., \& {Owen}, F.~N. 1995, \apj, 447, 582

\bibitem[{{Clark}(1980)}]{clark_b_1980}
{Clark}, B.~G. 1980, \aap, 89, 377

\bibitem[{{Cornwell}(2008)}]{cornwell_t_2008}
{Cornwell}, T.~J. 2008, IEEE Journal of Selected Topics in Signal Processing,
  2, 793

\bibitem[{Djamdji {et~al.}(1993)Djamdji, Bijaoui, \&
  Maniere}]{djamdji_geometrical_1993}
Djamdji, J.-P., Bijaoui, A., \& Maniere, R. 1993, in , 412--422

\bibitem[{{Doucet} \& {Gordon}(1999)}]{doucet_a_1999}
{Doucet}, A. \& {Gordon}, N.~J. 1999, in Society of Photo-Optical
  Instrumentation Engineers (SPIE) Conference Series, Vol. 3809, Signal and
  Data Processing of Small Targets 1999, ed. O.~E. {Drummond}, 241--255

\bibitem[{{Doucet} \& {Wang}(2005)}]{doucet_a_2005}
{Doucet}, A. \& {Wang}, X. 2005, IEEE Signal Processing Magazine, 22, 152

\bibitem[{{Fomalont}(1999)}]{fomalont_e_1999}
{Fomalont}, E.~B. 1999, in Astronomical Society of the Pacific Conference
  Series, Vol. 180, Synthesis Imaging in Radio Astronomy II, ed. G.~B.
  {Taylor}, C.~L. {Carilli}, \& R.~A. {Perley}, 301

\bibitem[{{Fromm} {et~al.}(2013){Fromm}, {Ros}, {Perucho}, {Savolainen},
  {Kadler}, {Lobanov}, \& {Zensus}}]{fromm_c_2013}
{Fromm}, C.~M., {Ros}, E., {Perucho}, M., {et~al.} 2013, \aap, 557, A105

\bibitem[{Giachetti(2000)}]{giachetti_matching_2000}
Giachetti, A. 2000, Image and Vision Computing, 18, 247

\bibitem[{{Hardee} {et~al.}(2005){Hardee}, {Walker}, \&
  {G{\'o}mez}}]{hardee_p_2005}
{Hardee}, P.~E., {Walker}, R.~C., \& {G{\'o}mez}, J.~L. 2005, \apj, 620, 646

\bibitem[{{H{\"o}gbom}(1974)}]{hogbom_j_1974}
{H{\"o}gbom}, J.~A. 1974, \aaps, 15, 417

\bibitem[{Holschneider {et~al.}(1989)Holschneider, Kronland-Martinet, Morlet,
  \& Tchamitchian}]{holschneider_a_1989}
Holschneider, M., Kronland-Martinet, R., Morlet, J., \& Tchamitchian, P. 1989,
  in Wavelets. Time-Frequency Methods and Phase Space, ed. J.-M. {Combes},
  A.~{Grossmann}, \& P.~{Tchamitchian}, 286

\bibitem[{{Hummel} {et~al.}(1992){Hummel}, {Muxlow}, {Krichbaum}, {Schalinski},
  \& {Witzel}}]{hummel_c_1992}
{Hummel}, C.~A., {Muxlow}, T.~W.~B., {Krichbaum}, T.~P.~and{Quirrenbach}, A.,
  {Schalinski}, C.~J., \& {Witzel}, A.~and{Johnston}, K.~J. 1992, \aap, 266, 93

\bibitem[{{Kovalev} {et~al.}(2008){Kovalev}, {Lobanov}, \&
  {Pushkarev}}]{kovalev_y_2008}
{Kovalev}, Y.~Y., {Lobanov}, A.~P., \& {Pushkarev}, A.~B.~and{Zensus}, J.~A.
  2008, \aap, 483, 759

\bibitem[{Lal {et~al.}(2010)Lal, Lobanov, \&
  Jiménez-Monferrer}]{lal_array_2010}
Lal, D.~V., Lobanov, A.~P., \& Jiménez-Monferrer, S. 2010, arXiv:1001.1477
  [astro-ph]

\bibitem[{Lewis(1995)}]{lewis_fast_1995}
Lewis, J. 1995, Vision Interface, 10, 120

\bibitem[{{Lister} {et~al.}(2009){Lister}, {Aller}, {Aller}, {Cohen}, {Kadler},
  {Kellermann}, {Ros}, {Savolainen}, \& {Zensus}}]{lister_m_2009}
{Lister}, M.~L., {Aller}, H.~D., {Aller}, M.~F., {et~al.} 2009, \aj, 137, 3718

\bibitem[{Lister {et~al.}(2013)Lister, Aller, Aller, Homan, Kellermann,
  Kovalev, Pushkarev, Richards, Ros, \& Savolainen}]{lister_mojave_2013}
Lister, M.~L., Aller, M.~F., Aller, H.~D., {et~al.} 2013, The Astronomical
  Journal, 146, 120

\bibitem[{{Lobanov} {et~al.}(2003){Lobanov}, {Hardee}, \&
  {Eilek}}]{lobanov_a_2003}
{Lobanov}, A., {Hardee}, P., \& {Eilek}, J. 2003, \nar, 47, 629

\bibitem[{{Lobanov}(1998{\natexlab{a}})}]{lobanov_a_1998c}
{Lobanov}, A.~P. 1998{\natexlab{a}}, \aaps, 132, 261

\bibitem[{{Lobanov}(1998{\natexlab{b}})}]{lobanov_a_1998}
{Lobanov}, A.~P. 1998{\natexlab{b}}, \aap, 330, 79

\bibitem[{{Lobanov}(2012)}]{lobanov_a_2012}
{Lobanov}, A.~P. 2012, in Square Kilometre Array: Paving the Way for the New
  21st Century Radio Astronomy Paradigm, ed. D.~{Barbosa}, S.~{Anton},
  L.~{Gurvits}, \& D.~{Maia} (Springer-Verlag: Belrin Heidelrberg), 75

\bibitem[{{Lobanov} \& {Zensus}(1999)}]{lobanov_a_1999}
{Lobanov}, A.~P. \& {Zensus}, J.~A. 1999, \apj, 521, 509

\bibitem[{{Lobanov} \& {Zensus}(2001)}]{lobanov_a_2001}
{Lobanov}, A.~P. \& {Zensus}, J.~A. 2001, Science, 294, 128

\bibitem[{Mallat(1989)}]{mallat_a_1989}
Mallat, S.~G. 1989, IEEE Transactions on Pattern Analysis and Machine
  Intelligence, 11, 674

\bibitem[{Men'shchikov {et~al.}(2012)Men'shchikov, André, Didelon, Motte,
  Hennemann, \& Schneider}]{shchikov_multiscale_2012}
Men'shchikov, A., André, P., Didelon, P., {et~al.} 2012, Astronomy and
  Astrophysics, 542, A81

\bibitem[{Murtagh {et~al.}(1995)Murtagh, Starck, \&
  Bijaoui}]{murtagh_image_1995}
Murtagh, F., Starck, J.-L., \& Bijaoui, A. 1995, Astronomy and Astrophysics
  Supplement Series, 112, 179

\bibitem[{Myint {et~al.}(2008)Myint, Yuan, Cerveny, \&
  Giri}]{myint_comparison_2008}
Myint, S.~W., Yuan, M., Cerveny, R.~S., \& Giri, C.~P. 2008, Sensors, 8, 1128

\bibitem[{Pan {et~al.}(2010)Pan, Xie, \& Wang}]{pan_equivalence_2010}
Pan, B., Xie, H., \& Wang, Z. 2010, Applied Optics, 49, 5501

\bibitem[{Paulson {et~al.}(2010)Paulson, Ezekiel, \&
  Wu}]{paulson_waveletbased_2010}
Paulson, C., Ezekiel, S., \& Wu, D. 2010, in , 77040M--77040M--12

\bibitem[{{Pearson}(1999)}]{pearson_t_1999}
{Pearson}, T.~J. 1999, in Astronomical Society of the Pacific Conference
  Series, Vol. 180, Synthesis Imaging in Radio Astronomy II, ed. G.~B.
  {Taylor}, C.~L. {Carilli}, \& R.~A. {Perley}, 335

\bibitem[{{Pushkarev} {et~al.}(2012){Pushkarev}, {Hovatta}, {Kovalev},
  {Lobanov}, \& {Savolainen}}]{pushkarev_a_2012}
{Pushkarev}, A.~B., {Hovatta}, T., {Kovalev}, Y.~Y.~and{Lister}, M.~L.,
  {Lobanov}, A.~P., \& {Savolainen}, T.~and{Zensus}, J.~A. 2012, \aap, 545,
  A113

\bibitem[{Rué \& Bijaoui(1997)}]{rue_a_1997}
Rué, F. \& Bijaoui, A. 1997, Experimental Astronomy, 7, 129

\bibitem[{Seymour \& Widrow(2002)}]{seymour_multiresolution_2002}
Seymour, M.~D. \& Widrow, L.~M. 2002, The Astrophysical Journal, 578, 689

\bibitem[{Shensa(1992)}]{shensa_the_1992}
Shensa, M.~J. 1992, IEEE Transactions on Signal Processing, 40, 2464

\bibitem[{{Sidenbladh} {et~al.}(2004){Sidenbladh}, {Svenson}, \&
  {Schubert}}]{sidenbladh_h_2004}
{Sidenbladh}, H., {Svenson}, P., \& {Schubert}, J. 2004, in Society of
  Photo-Optical Instrumentation Engineers (SPIE) Conference Series, Vol. 5429,
  Signal Processing, Sensor Fusion, and Target Recognition XIII, ed.
  I.~{Kadar}, 306--314

\bibitem[{Starck \& Murtagh(2006)}]{starck_astronomical_2006}
Starck, J.-L. \& Murtagh, F. 2006, Astronomical image and data analysis
  (Springer)

\bibitem[{Unser(1999)}]{unser_splines_1999}
Unser, M. 1999, IEEE Signal Processing Magazine, 16, 22

\bibitem[{{Unwin} {et~al.}(1997){Unwin}, {Wehrle}, {Lobanov}, {Madejski}, \&
  {Aller}}]{unwin_s_1997}
{Unwin}, S.~C., {Wehrle}, A.~E., {Lobanov}, A.~P.~and{Zensus}, J.~A.,
  {Madejski}, G.~M., \& {Aller}, M.~F.~and{Aller}, H.~D. 1997, \apj, 480, 596

\bibitem[{Vanderbrug \& Rosenfeld(1977)}]{vanderbrug_twostage_1977}
Vanderbrug, G. \& Rosenfeld, A. 1977, IEEE Transactions on Computers, C-26, 384

\bibitem[{Walker {et~al.}(2008)Walker, Ly, Junor, \& Hardee}]{walker_vlba_2008}
Walker, R.~C., Ly, C., Junor, W., \& Hardee, P.~J. 2008, Journal of Physics:
  Conference Series, 131, 012053

\bibitem[{Witkin {et~al.}(1987)Witkin, Terzopoulos, \&
  Kass}]{witkin_signal_1987}
Witkin, A., Terzopoulos, D., \& Kass, M. 1987, International Journal of
  Computer Vision, 1, 133

\bibitem[{{Yuan} {et~al.}(1998){Yuan}, {Elvidge}, \& {Lunetta}}]{yuan_d_1998}
{Yuan}, D., {Elvidge}, C.~D., \& {Lunetta}, R. 1998, in Remote Sensing Change
  Detection, Environmental Monitoring Methods andApplications, ed. M.~{Eden} \&
  J.~{Parry} (Ann Arbor Press), 1

\bibitem[{{Zensus} {et~al.}(1995){Zensus}, {Cohen}, \& {Unwin}}]{zensus_j_1995}
{Zensus}, J.~A., {Cohen}, M.~H., \& {Unwin}, S.~C. 1995, \apj, 443, 35

\bibitem[{Zheng \& Chellappa(1993)}]{zheng_computational_1993}
Zheng, Q. \& Chellappa, R. 1993, IEEE Transactions on Image Processing, 2, 311

\bibitem[{Zitová \& Flusser(2003)}]{zitova_image_2003}
Zitová, B. \& Flusser, J. 2003, Image and Vision Computing, 21, 977

\end{thebibliography}

\end{document}